\documentclass[a4paper,11pt]{article}
\usepackage{jcappub} 
\usepackage{lineno}
\usepackage{multirow}

\newcommand{\kmax}{k_{\rm max}}
\newcommand{\hmpc}{h~{\rm Mpc}^{-1}}
\newcommand{\be}{\begin{equation}}
\newcommand{\ee}{\end{equation}}
\newcommand{\bfk}{\mbox{\boldmath$k$}}

\newcommand{\bfq}{\mbox{\boldmath$q$}}
\newcommand{\de}{\mathrm{d}}
\newcommand{\PL}{P_{\rm L}}
\newcommand{\gcdm}{$\gamma$CDM}
\newcommand{\gnucdm}{$\gamma \nu$CDM}

\input{journalabbvr.texinc}

\title{\boldmath Modified gravity and massive neutrinos: constraints
  from the full shape analysis of BOSS galaxies and forecasts for
  Stage IV surveys}

\author[a,b,c]{Chiara Moretti,}
\author[a]{Maria Tsedrik,}
\author[a]{Pedro Carrilho,}
\author[a,d]{and Alkistis Pourtsidou}

\affiliation[a]{Institute for Astronomy, The University of Edinburgh,
  Royal Observatory, Edinburgh EH9 3HJ, UK}

\affiliation[b]{SISSA - International School for Advanced Studies, Via
  Bonomea 265, 34136 Trieste, Italy}

\affiliation[c]{INAF – Osservatorio Astronomico di Trieste, Via
  Tiepolo 11, I-34143 - Trieste, Italy}

\affiliation[d]{Higgs Centre for Theoretical Physics, School of
  Physics and Astronomy, The University of Edinburgh, Edinburgh EH9
  3FD, UK}

\emailAdd{cmoretti@sissa.it, mtsedrik@ed.ac.uk,
  pedro.carrilho@ed.ac.uk, alkistis.pourtsidou@ed.ac.uk}

\abstract{ We constrain the growth index $\gamma$ by performing a
  full-shape analysis of the power spectrum multipoles measured from
  the BOSS DR12 data. We adopt a theoretical model based on the
  Effective Field theory of the Large Scale Structure (EFTofLSS) and
  focus on two different cosmologies: \gcdm{} and \gnucdm{}, where we
  also vary the total neutrino mass. We explore different choices for
  the priors on the primordial amplitude $A_s$ and spectral index
  $n_s$, finding that informative priors are necessary to alleviate
  degeneracies between the parameters and avoid strong projection
  effects in the posterior distributions.  Our tightest constraints
  are obtained with 3$\sigma$ Planck priors on $A_s$ and $n_s$: we
  obtain $\gamma = 0.647 \pm 0.085$ for \gcdm{} and $\gamma =
  0.612^{+0.075}_{-0.090}$, $M_\nu < 0.30$ for \gnucdm{} at 68\% c.l.,
  in both cases $\sim 1\sigma$ consistent with the $\Lambda$CDM
  prediction $\gamma \simeq 0.55$. Additionally, we produce forecasts
  for a Stage-IV spectroscopic galaxy survey, focusing on a DESI-like
  sample. We fit synthetic data-vectors for three different galaxy
  samples generated at three different redshift bins, both
  individually and jointly.  Focusing on the constraining power of the
  Large Scale Structure alone, we find that forthcoming data can give
  an improvement of up to $\sim 85\%$ in the measurement of $\gamma$
  with respect to the BOSS dataset when no CMB priors are imposed. On
  the other hand, we find the neutrino mass constraints to be only
  marginally better than the current ones, with future data able to
  put an upper limit of $M_\nu < 0.27~{\rm eV}$. This result can be
  improved with the inclusion of Planck priors on the primordial
  parameters, which yield $M_\nu < 0.18~{\rm eV}$.}

\begin{document}
\maketitle
\flushbottom

\section{Introduction}
\label{sec:intro}
The current concordance model of cosmology, the $\Lambda$CDM model,
has been confirmed over the last few decades by increasingly precise
observations, spanning from the Cosmic Microwave Background (CMB)
\cite{planck2018cosmo} to the Large Scale Structure (LSS)
\cite{eboss2021, kids2021, des2022}. In the standard framework, the
gravitational interaction is described by General Relativity (GR) and
the energy content of the Universe is composed of matter, which
includes cold dark matter (CDM) and baryons, and dark energy in the
form of a cosmological constant ($\Lambda$), introduced to explain the
observed accelerated expansion of the Universe \cite{riess1998,
  perlmutter1999}. Despite the many successes of the model at fitting
observations, major fundamental questions concerning the nature of the
dark components remain unanswered. Furthermore, the increasing
accuracy of recent observations has brought to light tensions in the
cosmological parameters when measured from early versus late time
probes \cite{verde2019, divalentino2021, perivolaropoulos2022}.  This
context offers the perfect breeding ground for alternative theories,
which can explain the accelerated expansion of the Universe without
the need for a dark energy component, or mitigate the tensions in
cosmological parameter measurements.

Confirming or disproving the standard picture is the primary goal of
ongoing and forthcoming Stage-IV galaxy redshift surveys, such as DESI
\cite{desi2016}, the Euclid satellite mission \cite{laureijs2011},
Rubin's Legacy Survey of Space and Time \cite{lsst2018} and the Nancy
Grace Roman space telescope \cite{wfirst2015}.  These will map the 3D
galaxy distribution with unprecedented accuracy over extremely large
volumes, delivering high-precision measurements of the cosmological
observables.  The latter will allow to measure the cosmological
parameters to sub-percent precision and disentangle between different
gravity models \cite{amendola2018, Alam:2020jdv}.  A crucial element
to achieve this goal is the availability of an accurate and reliable
theoretical model for the cosmic observables, able to describe the
nonlinear regime of structure formation.

The standard approach to describe the impact of nonlinear evolution on
clustering observables is based on Perturbation Theory (see
\cite{bernardeau2002} for a review), which features contributions to
the power spectrum in the form of convolution integrals.  If computed
with standard integration techniques, such integrals are too
computationally expensive to be evaluated over a large number of
points in parameter space; for this reason, previous analyses have
relied on a template-fitting approach.  However, recent advances on
the theoretical and numerical fronts have allowed for a full-shape
analysis of the summary statistics. On the one hand, the development
of the Effective Field Theory of the Large Scale Structure (EFTofLSS,
\cite{baumann2012, carrasco2012, pietroni2012}) has allowed to include
the impact of unknown small-scale physics on the intermediate scales,
thus extending the validity range of the model. On the other hand, the
development of FFTlog-based algorithms \cite{mcewen2016, fang2017,
  simonovic2018} has sped up the computation of the convolution
integrals by orders of magnitude, enabling sampling of the full
parameter space.

Constraints on the parameters of the $\Lambda$CDM model obtained from
full-shape fits of the clustering measurements of the Baryon
Oscillation Spectroscopic Survey (BOSS) have been presented in a
number of recent papers, see e.g. \cite{damico2020, ivanov2020b,
  chen2022}.  In terms of constraints on beyond-$\Lambda$CDM scenarios
from the same dataset, the BOSS collaboration adopted a
template-fitting approach \cite{beutler2014a, beutler2014b,
  mueller2018}.  Additionally, a full-shape analysis has been
performed in \cite{ivanov2020c, damico2021, semenaite2022,
  carrilho2023, simon2023, piga2023} in the context of specific
beyond-$\Lambda$CDM models.  In this work, we focus instead on a model
independent way of checking deviations from base $\Lambda$CDM, the so-called growth
index parameterisation \cite{Peebles1980,linder2005}.  This is a
one-parameter extension of the standard model which allows the growth
of structures to deviate from the prediction of $\Lambda$CDM by means
of the growth index, $\gamma$.  Despite the model being purely
phenomenological, a measurement of $\gamma$ inconsistent with its
$\Lambda$CDM prediction would provide a clear detection of deviations
from the standard cosmological model.  On the other hand, it has been shown that the growth index parameterisation can provide a good fit for specific modified gravity models, such as the DGP model \cite{Dvali:2000hr}, with $\gamma \simeq 0.67$ \cite{linder2007, gong2008}, and a class of Horndeski theories \cite{wen2023}, although in that case a time dependent growth index is required. For this reason a precise measurement of the growth index is
among the objectives of forthcoming Stage-IV surveys \cite{desi2016,
  amendola2018, euclidforecast2020}.  In addition to testing the
theory of gravity, a key science goal of forthcoming galaxy surveys is
a measurement of the neutrino mass.  It is well known that massive
neutrinos can suppress structure formation below a specific `free
streaming' scale, thus leaving a distinct signature on cosmological
observables (see \cite{lesgourgues2006} for a review).  However,
current LSS data alone are not sufficiently accurate to constrain the
neutrino mass, or even provide competitive upper bounds with respect
to CMB measurements.  This picture is expected to change with Stage-IV
measurements of the LSS, that promise to achieve sufficient precision
to allow for a detection of the neutrino mass.

In this work, we perform the full-shape analysis of the galaxy power
spectrum as measured from BOSS Data Release 12, providing joint
constraints for the growth index and the total neutrino mass.  The
paper is structured as follows: in Sec.~\ref{sec:theory} we give an
overview of the theoretical model for the power spectrum, and describe
the dataset and our analysis setup in Sec.~\ref{sec:data-setup}. Our
main results are presented in Sec.~\ref{sec:results}, where we also
explore the impact on the constraints of different prior choices for
the primordial parameters $A_s$ and $n_s$. Additionally, we present
forecasts on $\gamma$ and $M_\nu$ for Stage-IV spectroscopic surveys,
focusing on a DESI-like galaxy sample. Finally, we summarise our
conclusions in Sec.~\ref{sec:conclusion}.

%
\section{Theoretical model}
\label{sec:theory}
We model the nonlinear galaxy power spectrum in redshift space using
the Effective Field Theory of the Large Scale Structure (EFTofLSS,
\cite{baumann2012, carrasco2012, perko2016, delabella2017}), which
allows to model the effects of unknown small-scale physics on mildly
nonlinear scales with the introduction of scale-dependent terms in the
theoretical model. In particular, we follow a prescription analogous
to the one of \cite{chudaykin2020}, but employ an independent
implementation \cite{oddo2020, oddo2021, rizzo2023},
whose redshift space modelling for the power spectrum has been
validated on N-body simulations in \cite{carrilho2021, tsedrik2023}.
The model features several ingredients, including loop corrections,
EFTofLSS counterterms, a bias prescription and an infrared resummation
routine; we outline these below, but refer to \cite{carrilho2023} for
a more detailed description.
\subsection{Nonlinear power spectrum}
\label{sec:nl-pk}
The bias expansion we adopt is based on a perturbative expansion of
the galaxy density field $\delta_g$ \cite{mcdonald2009, assassi2014,
  desjacques2018, abidi2018}, which includes contributions from the
underlying matter density field and large-scale tidal fields. Here we
only consider terms up to third order in perturbations of $\delta$:
\be \delta_g = b_1 \delta + \frac{b_2}{2} \delta^2 + b_{\mathcal
  G_2} \mathcal{G}_2 + b_{\Gamma_3} \Gamma_3 + \epsilon \, ,
\label{eq:bias} \ee
where $b_1$ and $b_2$ are respectively the linear and quadratic local
bias parameters, $\mathcal{G}_2$ and $\Gamma_3$ are non-local
functions of the density field and $\epsilon$ is a stochastic
contribution.  The one loop anisotropic galaxy power spectrum can then
be written as:
\begin{align}
P_{gg}(\bfk) = &Z_1^2(\bfk) \PL(k) + 2 \int \de^3 \bfq \, \left[
  Z_2(\bfq, \bfk - \bfq) \right]^2\PL(q) \PL(\vert \bfk - \bfq \vert)
\nonumber \\ &+ 6 Z_1(\bfk)\PL(k) \int \de^3 \bfq \, Z_3(\bfk, \bfq,
-\bfq)\PL(q) + P_{\rm ctr}(\bfk)
+P_{\epsilon\epsilon}(\bfk) \, ,
\label{eq:looppk}
\end{align}
where $\PL$ is the linear power spectrum\footnote{For all analyses
performed here we consider a massive neutrino component, even when we
do not vary the total neutrino mass as a free parameter. Motivated by
\cite{castorina2014, castorina2015}, we always use the cold dark matter + baryon as
linear power spectrum in Eq.~\ref{eq:looppk} and following equations.}
and $Z_1(\bfk)$, $Z_2(\bfk_1, \bfk_2)$, $Z_3(\bfk_1, \bfk_2, \bfk_3)$
are the redshift-space kernels \cite{scoccimarro1999}, whose full
expressions for the bias basis of Eq.~\ref{eq:bias} can be found in
Appendix A of \cite{ivanov2020b}.  The integrals in
Eq.~\ref{eq:looppk} are computed under the assumption that time and
scale evolution are fully separable, i.e. we adopt the so-called
Einstein-de Sitter approximation. The latter was shown to be better
than 1\% accurate in $\Lambda$CDM at the redshift of BOSS
\cite{donath2020}, and allows us to compute the integrals only once
per set of cosmological parameters, and then re-scale them using the
linear growth factor to get the observables at the desired redshift.

The power spectrum of Eq.~\ref{eq:looppk} includes two additional
contributions: the EFTofLSS counterterms, and noise. The EFTofLSS
counterterms can be written as:
\begin{align}
P_{\rm ctr}(k, \mu) = &- 2 \tilde{c}_0 k^2 P_L(k) - 2 \tilde{c}_2 k^2 f
\mu^2 P_L(k) - 2 \tilde{c}_4 k^2 f^2 \mu^4 P_L(k)
\nonumber\\ &+c_{\nabla^4 \delta} k^4 f^4 \mu^4 (b_1 + f \mu^2)^2
P_L(k)\,,
\label{eq:couterterms}
\end{align}
where $\mu$ is the cosine of the angle between the wavevector $\bfk$
and the line of sight.  Following \cite{chudaykin2020}, we re-define
the $k^2$-counterterm parameters in order to have separate
contributions to each multipole.  We model the noise as:
\be P_{\epsilon \epsilon}(k, \mu) = N + e_0 k^2 + e_2 k^2\mu^2
\, ,
\label{eq:noise}
\ee
where $N$ is a constant that includes deviations from pure Poisson
shot noise, and we have two additional scale-dependent terms.  We
account for the nonlinear evolution of the BAO peak \cite{crocce2008,
  senatore2015, vlah2016} by means of an infrared resummation routine,
applied by splitting the linear power spectrum in a smooth broadband
component (computed using the fit of \cite{eisenstein1998}) and a
wiggle part. Full details of the approach we use can be found in
Sec.~2.2.2 of \cite{carrilho2023}.  In total we have a set of 11
nuisance parameters {\it per redshift bin}:
\be \theta = \{ b_1,~b_2,~b_{\mathcal{G}_2},~b_{\Gamma_3},
~N,~e_0,~e_2,~c_0,~c_2,~c_4,~c_{\nabla^4 \delta}\} \, . \ee

In our baseline analysis we keep the total neutrino mass fixed to its
fiducial value $M_{\nu} = 0.06~{\rm eV}$. We model the scale-dependent
suppression induced by massive neutrinos by computing the cdm+baryon
linear power spectrum and using it as input for the theoretical model
(instead of the total matter one, which also includes the neutrino
contribution).  Additionally, we consider a cosmology with free
neutrino mass, and we modify the theoretical model as described in
more detail in Sec.~\ref{sec:neutrinos}.

In order to include the impact of the fiducial cosmology assumed when
converting redshifts to distances in the data, we correct wavenumbers
and angles by applying Alcock-Paczynski distortions \cite{alcock1979}:
\begin{align}
    \bar{k}^2=k^2\left(\frac{H_0^{\rm
        fid}}{H_{0}}\right)^2\left(\left(\frac{H}{H^{\rm
        fid}}\right)^2 \mu^2+\left(\frac{D_{A}^{\rm
        fid}}{D_A}\right)^2(1-\mu^2)\right)\,,\\ \bar{\mu}^2=\mu^2\left(\frac{H}{H^{\rm
        fid}}\right)^2\left(\left(\frac{H}{H^{\rm fid}}\right)^2
    \mu^2+\left(\frac{D_{A}^{\rm
        fid}}{D_A}\right)^2(1-\mu^2)\right)^{-1}\,,
\end{align}
as well as re-scaling the power spectrum by a factor
\be A_{\rm AP}=\left(\frac{H_0^{\rm
    fid}}{H_{0}}\right)^3\frac{H}{H^{\rm fid}}\left(\frac{D_{A}^{\rm
    fid}}{D_A}\right)^2\, , \ee
where $H_0$ is the Hubble factor today, $D_A$ is the angular diameter
distance, and $^{\rm fid}$ refers to quantities evaluated in the
fiducial cosmology. Finally, we project the anisotropic power spectrum
to multipoles in the usual way:
\be P_\ell(k)=\frac{2\ell+1}{2}\int_{-1}^1{\rm d} \mu\,
P\left(\bar{k}(k,\mu),\bar{\mu}(k,\mu)\right)\mathcal{P}_\ell(\mu)\,,
\ee
where $\mathcal{P}_\ell(\mu)$ is the Legendre polynomial of order $\ell$. 

Our implementation allows for bypassing the
\texttt{camb}\footnote{\url{https://camb.info/}} Boltzmann solver
\citep{Lewis:1999bs} in order to use linear power spectrum emulators,
namely
\texttt{bacco}\footnote{\url{https://baccoemu.readthedocs.io/en/latest/}}
\cite{arico2021} and
\texttt{CosmoPower}\footnote{\url{https://alessiospuriomancini.github.io/cosmopower/}}
\cite{SpurioMancini:2021ppk}. We note that this speeds up the model
evaluations by roughly two orders of magnitude.
\subsection{The growth index \texorpdfstring{$\gamma$}{} parameterisation}
\label{sec:gamma}
In order to check for deviations from $\Lambda$CDM, we adopt the
phenomenological parameterisation proposed in
\cite{Peebles1980,linder2005}. Specifically, we compute the growth
rate $f = {\rm d} \ln D / {\rm d} \ln a$ as:
\be f_\gamma(a) = \Omega_m(a)^\gamma \, ,
    \label{eq:growth-rate-gamma}
\ee
where $\Omega_m(a)=\Omega_{m,0} a^{-3} / E^2(a)$ is the time dependent
matter density parameter, with $E(a)$ the dimensionless Hubble
factor. Eq.~\ref{eq:growth-rate-gamma} has been shown to be $0.2\%$
accurate for $\Lambda$CDM scenarios, with $\gamma=0.55$
\cite{linder2005}.  We compute the linear growth factor $D$ by
numerically integrating Eq.~\ref{eq:growth-rate-gamma}:
\be D_\gamma(a) = \exp \left[ - \int {\rm d}a'~\frac{f_\gamma(a')}{a'}
  \right] \, .
    \label{eq:lin-growth-gamma}
\ee
We normalise the growth rate to its $\Lambda$CDM value at high
redshift, i.e. during matter domination.  The growth functions from
Eq.~\ref{eq:growth-rate-gamma} and \ref{eq:lin-growth-gamma} are then
used to re-scale the linear power spectrum computed at redshift $z=0$
and construct the redshift space multipoles as described in
Sec.~\ref{sec:nl-pk}.  An analysis of the behaviour of $\gamma$ over cosmic history can be found in \cite{calderon2019, calderon2020}, while \cite{wen2023} propose an alternative parameterisation, that extends $\gamma$ to be a function of redshift.
\subsection{Massive neutrinos}
\label{sec:neutrinos}
The scale-dependent suppression in growth induced by massive neutrinos
free streaming (see \cite{lesgourgues2006} for a review) is modelled
following the prescription of \cite{castorina2014}, which consists in
using the cold dark matter + baryon power spectrum $P_{cb}$ instead of
the total matter $P_m$ to compute the galaxy power spectrum
multipoles. While not exact, this approach was shown to be
sufficiently accurate for survey volumes similar to that of BOSS in
\cite{noriega2022}, especially for small neutrino masses.  For our
baseline analysis with fixed $M_{\nu}=0.06~{\rm eV}$ we modify the
model described in Sec.~\ref{sec:nl-pk} by substituting $\delta$ with
$\delta_{cb}$ in Eq.~\ref{eq:bias}, and thus $P_L$ with $P_{cb}$ in
what follows.

For the case where the neutrino mass is varied as a free parameter,
given the bigger impact that neutrino masses larger than the minimal
one have on the power spectrum, we perform the following modifications
to the model:
\begin{itemize}
\item we compute the linear $P_{cb}$ twice, once for each effective
  redshift of the galaxy sample, in order to properly model the
  scale-dependent suppression introduced by massive neutrinos;
\item we modify our infrared resummation routine, since the
  Eisenstein-Hu prescription we adopt for the broadband power spectrum
  does not include a dependence on massive neutrinos.  We use instead
  a discrete sine transform approach to perform the wiggle-no wiggle
  decomposition \cite{baumann2018, giblin2019};
\item we compute the growth rate of Eq.~\ref{eq:growth-rate-gamma}
  using only the cdm+baryon density $\Omega_{cb}$ \cite{boyle2021};
\item the resulting growth factor is normalised to its $\Lambda$CDM
  counterpart at the effective redshift, so that differences in the
  amplitude only come from the $\gamma$ parameter.
\end{itemize}

In principle, a more accurate modelling of the impact of massive neutrinos
would require a modification of the perturbation theory kernels and
the use of a scale dependent growth rate (see e.g. \cite{saito2009,
  levi2016, senatore2017, aviles2020, aviles2021}).  However, we do not
expect any relevant effect for the case of BOSS measurements given the
size of the error bars \cite{noriega2022}.
%
%
\section{Data and analysis setup}
\label{sec:data-setup}
Our main analysis setup and dataset follow the ones described in
\cite{carrilho2023}. We summarise here the main points, but refer to
that work for more details.
\subsection{Dataset}
\label{sec:data}
We use the galaxy power spectrum multipoles of BOSS DR12
\cite{alam2015, gilmarin2016, beutler2017}, which is split in two
galaxy samples (CMASS and LOWZ) and covers two different sky cuts (NGC
and SGC). The measurements are performed by splitting the sample into
two redshift bins, with effective redshift $z_1=0.38$ and $z_3=0.61$
and widths $\Delta z = 0.3$ and $0.25$, respectively.  In particular,
we use the windowless measurements provided in \cite{philcox2022},
based on the windowless estimator of \cite{philcox2021a,
  philcox2021b}.  The power spectrum measurements are complemented
with BAO measurements of the Alcock-Paczynski parameters
$\alpha_{\perp}$, $\alpha_{\parallel}$ obtained from the same BOSS
data release and also provided by \cite{philcox2022}, as well as
pre-reconstruction BAO measurements at low redshift (6DF survey
\cite{beutler2011} and SDSS DR7 MGS \cite{ross2015}) and high redshift
measurements of the Hubble factor and angular diameter distance from
the Ly-$\alpha$ forest auto and cross-correlation with quasars from
eBOSS DR12 \cite{desbourboux2020}.  We use a numerical covariance
matrix computed from 2048 `MultiDark-Patchy' mock catalogs
\cite{kitaura2016, rodrigueztorres2016}, also provided in its
windowless form by \cite{philcox2022}.
\subsection{Analysis setup}
\label{sec:setup}
We fit all three multipoles of the galaxy power spectrum up to $\kmax
= 0.2~\hmpc$. The nonlinear power spectrum is modelled with the
EFTofLSS as described in Sec.~\ref{sec:nl-pk}, with a custom
implementation which takes advantage of the
\texttt{FAST-PT}\footnote{\url{https://github.com/JoeMcEwen/FAST-PT}}
algorithm \cite{mcewen2016, fang2017} to compute the convolution
integrals of Eq.~\ref{eq:looppk} in $\mathcal{O}(10^{-2})$
seconds. The theoretical prediction is then combined with an
independently developed likelihood pipeline, where we sample the
parameter space by means of the affine-invariant sampler implemented
in the {\tt
  emcee}\footnote{\url{https://emcee.readthedocs.io/en/stable/}}
package \cite{foremanmackey2013}, using the integrated
auto-correlation time\footnote{In particular, we compute $\tau$ every 100 steps of the chain and check that, for each parameter, two conditions are satisfied: $\tau < {\rm nstep}/50$ and $\Delta \tau / \tau < 0.01$, with $\Delta \tau$ the difference between the current value of $\tau$ and its value at the previous check.} as convergence diagnostic for our MCMC chains
\cite{goodman2010}. For the Stage-IV forecasts presented in
Sec.~\ref{sec:forecasts} we use the pre-conditioned MonteCarlo method
implemented in {\tt
  pocomc}\footnote{\url{https://pocomc.readthedocs.io/en/latest/}}
\cite{karamanis2022a, karamanis2022b}, which allows for a speed up in
the sampling of parameters. We check the two samplers give consistent
results.

We marginalise analytically over the nuisance parameters that enter
the model linearly \cite{damico2020, damico2021}, namely the
EFT counterterm parameters, the noise parameters, and $b_{\Gamma 3}$. The
parameter space is thus restricted to 18 free parameters (19 when we
also vary the total neutrino mass):
\be \{ \omega_c,~\omega_b,~h,~n_s,~A_s,~\gamma,~(M_{\nu});~{\bf
  b_1},~{\bf b_2},~{\bf b_{\mathcal{G}_2}}\}\, , \ee
where the bias parameters assume different values in each redshift bin
and sky cut (e.g. ${\bf b_1} = \left[ b_1^{\rm NGC,z1},~b_1^{\rm
    SGC,z1},~b_1^{\rm NGC,z3},~b_1^{\rm SGC,z3} \right]$).

Concerning priors\footnote{We denote the uniform distribution with
edges $a,b$ as $\mathcal{U}(a,b)$, and the normal distribution with
mean $a$ and standard deviation $b$ as $\mathcal{N}(a,b)$.}, we impose
a Gaussian BBN prior on $\omega_b$\footnote{We use the results of
\cite{aver2015, cooke2018, schoneberg2019}, so that $\omega_b \in
\mathcal{N}(0.02268, 0.00038)$}, and broad, flat priors on the
cosmological parameters, matching the ranges of parameters allowed by
the \texttt{bacco} linear emulator:
\be \Omega_{cb} \in \mathcal{U}(0.06,0.7)\,,\ h \in
\mathcal{U}(0.5,0.9)\,, M_{\nu} \in \mathcal{U}(0,1) \, , \ee
where and $\Omega_{cb}=\Omega_c+\Omega_b$. We impose no prior
on the other cosmological parameters in our baseline analysis, but we
also explore the impact of including Planck priors on the primordial
parameters $A_s$ and $n_s$. In particular, for each cosmology explored
(\gcdm{}, \gnucdm{}), we have three options:
\begin{itemize}
\item one with no priors on $A_s$, $n_s$ (labelled {\it baseline});
\item one with a $3\sigma$ Gaussian Planck prior on $A_s$:
  $\ln(10^{10}~A_s) \in \mathcal{N}(3.044, 0.042)$ (labelled {\it  prior $A_s$});
  \item and one with a $3\sigma$ Gaussian Planck prior on both $A_s$
    and $n_s$: $\ln(10^{10}~A_s) \in \mathcal{N}(3.044, 0.042)$ and
    $n_s \in \mathcal{N}(0.9649, 0.0126)$ (labelled {\it prior
      $A_s$,$n_s$}).
\end{itemize}
For both parameters, the Gaussian priors are centered on the best-fit values from \cite{planck2018cosmo} and have width corresponding to the 3$\sigma$ error from the same work.
For the nuisance parameters, we adopt those of \cite{philcox2022}:
\begin{gather}
    b_1\in
    \mathcal{U}(0,4)\,,\ b_2\in\mathcal{N}(0,1)\,,\ b_{\mathcal{G}_2}\in\mathcal{N}(0,1)\,,\ b_{\Gamma_3}\in\mathcal{N}\left(\frac{23}{42}(b_1-1),1\right)\,,\nonumber\\ N\in\mathcal{N}\left(\frac{1}{\bar{n}},\frac{2}{\bar{n}}\right)\,,\ e_0\in\mathcal{N}\left(0,\frac{2}{\bar{n}k_{\rm
        NL}^2}\right)\,,\ e_2\in\mathcal{N}\left(0,\frac{2}{\bar{n}k_{\rm
        NL}^2}\right)\,,\ \frac{c_0}{[{\rm
          Mpc}/h]^2}\in\mathcal{N}(0,30)\,,\label{eq:prior_base}\\ \frac{c_2}{[{\rm
          Mpc}/h]^2}\in\mathcal{N}(30,30)\,,\ \frac{c_4}{[{\rm
          Mpc}/h]^2}\in\mathcal{N}(0,30)\,,\ \frac{c_{\nabla^4
        \delta}}{[{\rm Mpc}/h]^4}\in\mathcal{N}(500,500)\,.\nonumber
\end{gather}
%
%
\section{Results}
\label{sec:results}
\subsection{BOSS analysis}
\label{sec:boss-results}
\begin{figure}
    \centering
    \includegraphics[width=\columnwidth]{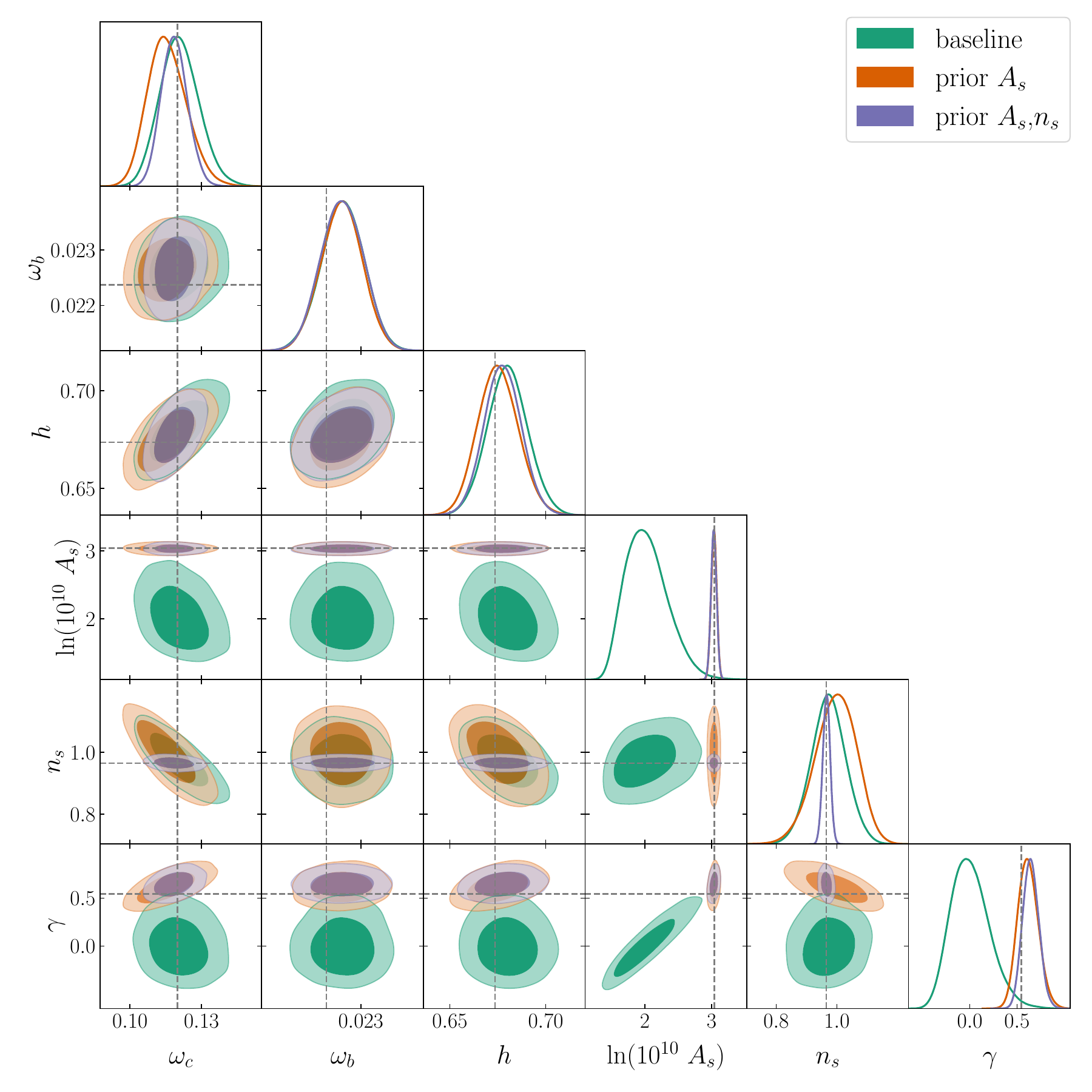}
    \caption{Marginalised posterior distribution for the cosmological
      parameters for the \gcdm{} cosmology and the three prior choices,
      as detailed in the legend. We fit all three multipoles and use
      $\kmax = 0.2~\hmpc$. Grey dashed lines mark the
      Planck best-fit values ($\Lambda$CDM prediction for $\gamma$).}
    \label{fig:gamma-cosmo}
\end{figure}

Here we present the main results of the analysis of the galaxy power
spectrum multipoles measured from the BOSS DR12; we cover the two
extensions of the base $\Lambda$CDM scenario discussed in the previous
sections: \gcdm{} and \gnucdm{}. For each cosmology, we explore the three
options for priors described in Sec.~\ref{sec:setup}.  We show the
posterior distributions for the \gcdm{} case with fixed neutrino mass
$M_{\nu}=0.06~{\rm eV}$ in Fig.~\ref{fig:gamma-cosmo} and summarise
the 68\% c.l., mean and best-fit values for the cosmological
parameters in Table~\ref{tab:gamma-bestfits}.  Plots for the full
parameter space can be found in Appendix~\ref{app:full-contours}. We
obtain $\gamma = 0.007^{+0.170}_{-0.229}~(0.149)$ for the baseline
priors, shown by green contours, $\gamma =
0.617^{+0.098}_{-0.110}~(0.614)$ (prior on $A_s$, orange contours) and
$\gamma = 0.647\pm 0.085~(0.655)$ (prior on $A_s$ and $n_s$, purple
contours), where values in parentheses refer to the best-fit values,
obtained as the maximum of the analytically marginalised posterior.

We first note that our baseline analysis is affected by strong
projection effects (also known as prior volume effects), in particular
involving the parameters regulating the amplitude of the power
spectrum, $A_s$ and $\gamma$: when no CMB priors are imposed, the
two-dimensional marginalised posteriors are shifted towards extremely
low values of $A_s$ and $\gamma$, along the degeneracy between the
two.  In fact, lowering the primordial amplitude $A_s$ or introducing
a low value for $\gamma$, which effectively increases the growth
factor, result in two opposite effects which can balance out and give
the same power spectrum.  An analogous behaviour for extensions of the
standard model was highlighted in \cite{carrilho2023}, especially for
the case of exotic (interacting) dark energy cosmology (see their
Fig.~7). Other works focusing on the full-shape analysis of the same
dataset also highlighted issues when trying to constrain extended
parameter models \cite{simon2022, piga2023}.  In particular in
\cite{piga2023} the authors provide constraints on nDGP modified
gravity \cite{Dvali:2000hr} using a similar EFTofLSS-based model, and
discuss the presence of projections due to the degeneracy between the
primordial amplitude $A_s$ and the nDGP $\Omega_{\rm rc}$
parameter. We study the presence of prior volume effects in more
detail in Sec.~\ref{sec:projection-effects} by generating and fitting
synthetic data, adopting the same covariance matrix used for our main
BOSS analysis, and by profiling the posterior.

Imposing a prior on $A_s$ breaks these degeneracies and shifts the two-dimensional
marginalised posterior closer to the Planck best-fit and $\Lambda$CDM
values, as can be seen in the orange contours of
Fig.~\ref{fig:gamma-cosmo}. In fact, the other cosmological parameters
are mostly unaffected, but the error on $\gamma$ is reduced by $\sim
48\%$. Moreover, adopting a prior on $n_s$ yields an additional $\sim
18\%$ improvement on $\gamma$.  Both the mean and best-fit values
obtained in these cases are slightly larger than the $\Lambda$CDM
prediction of $\gamma=0.55$, though still $\sim 1\sigma$ consistent. This
slight deviation is likely equivalent to the finding in previous full-shape BOSS analyses of a low value of $A_s$\footnote{This low amplitude is possibly arising due to prior volume/projection effects as seen in Refs.~\cite{carrilho2023, simon2022, damico2022}. We explore some of these effects in Sec.~\ref{sec:projection-effects}, but leave a more complete analysis for future work.} \cite{ivanov2020b, damico2020, carrilho2023} 
($\ln(10^{10}~A_s) \simeq 2.8$ as opposed to the Planck value
$\ln(10^{10}~A_s) = 3.044$): when we impose a prior on $A_s$ the model
tries to compensate by lowering the amplitude with a larger value for
$\gamma$. We also notice that the degeneracy with $A_s$ was masking a
degeneracy between $\gamma$, $\omega_c$ and $n_s$, which emerges when we impose a prior on $A_s$ and is then
broken when we apply a Planck prior on $n_s$.
\begin{table}[ht!]
\footnotesize
  \centering
  \begin{tabular} { l  c  c  c}
    Parameter &  baseline & prior on $A_s$ & prior on $A_s$, $n_s$\\
    \hline
	\hline
    \multirow{2}{2em}{$\omega_c       $}  & $0.1208^{+0.0076}_{-0.0086}$ & $0.1157^{+0.0072}_{-0.0090}$ & $0.1186^{+0.0052}_{-0.0058}$\\
    & (0.1216)	&   (0.1165)  & (0.1196) \\
    \hline
    \multirow{2}{2em}{$\omega_b       $}  & $0.02267\pm 0.00039        $ & $0.02265\pm 0.00038        $ & $0.02266\pm 0.00038        $\\
    & (0.02298)	&   (0.02292) & (0.02275)\\
    \hline
    \multirow{2}{2em}{$h              $}  & $0.680\pm 0.010            $ & $0.675\pm 0.011            $ & $0.678\pm 0.010          $\\
    & (0.682)	   	&   (0.682)	  & (0.681)  \\
    \hline
    \multirow{2}{2em}{$\ln(10^{10}~A_s)$} & $2.030^{+0.248}_{-0.371}      $ & $3.034\pm 0.042            $ & $3.033\pm 0.042            $\\
    & (2.190)	   	&   (3.055)	  & (3.065)  \\
    \hline
    \multirow{2}{2em}{$n_s            $}  & $0.973\pm 0.057            $ & $0.996^{+0.072}_{-0.062}   $ & $0.966\pm 0.012            $\\
    & (0.970)	   	&   (0.994)	  & (0.970)  \\
    \hline
    \multirow{2}{2em}{$\gamma         $}  & $0.007^{+0.170}_{-0.229}      $ & $0.617^{+0.098}_{-0.110}    $ & $0.647\pm 0.085            $\\
    & (0.149)	   	&   (0.614)	  & (0.655)  \\
    \hline
    \multirow{2}{2em}{$\sigma_8       $}  & $0.498^{+0.047}_{-0.096}   $ & $0.796\pm 0.030            $ & $0.801\pm 0.029            $\\
    & (0.534)	   	&   (0.807)	  & (0.819)  \\
    \hline
  \end{tabular}

  \caption{Mean values and 68\% c.l. values for \gcdm{} with fixed
    neutrino mass $M_{\nu}=0.06~{\rm eV}$ for the three prior
    choices. We show the best-fit values in parentheses, and include
    derived constraints on $\sigma_8$.}
  \label{tab:gamma-bestfits}
\end{table}

\begin{figure}
    \centering
    \includegraphics[width=\columnwidth]{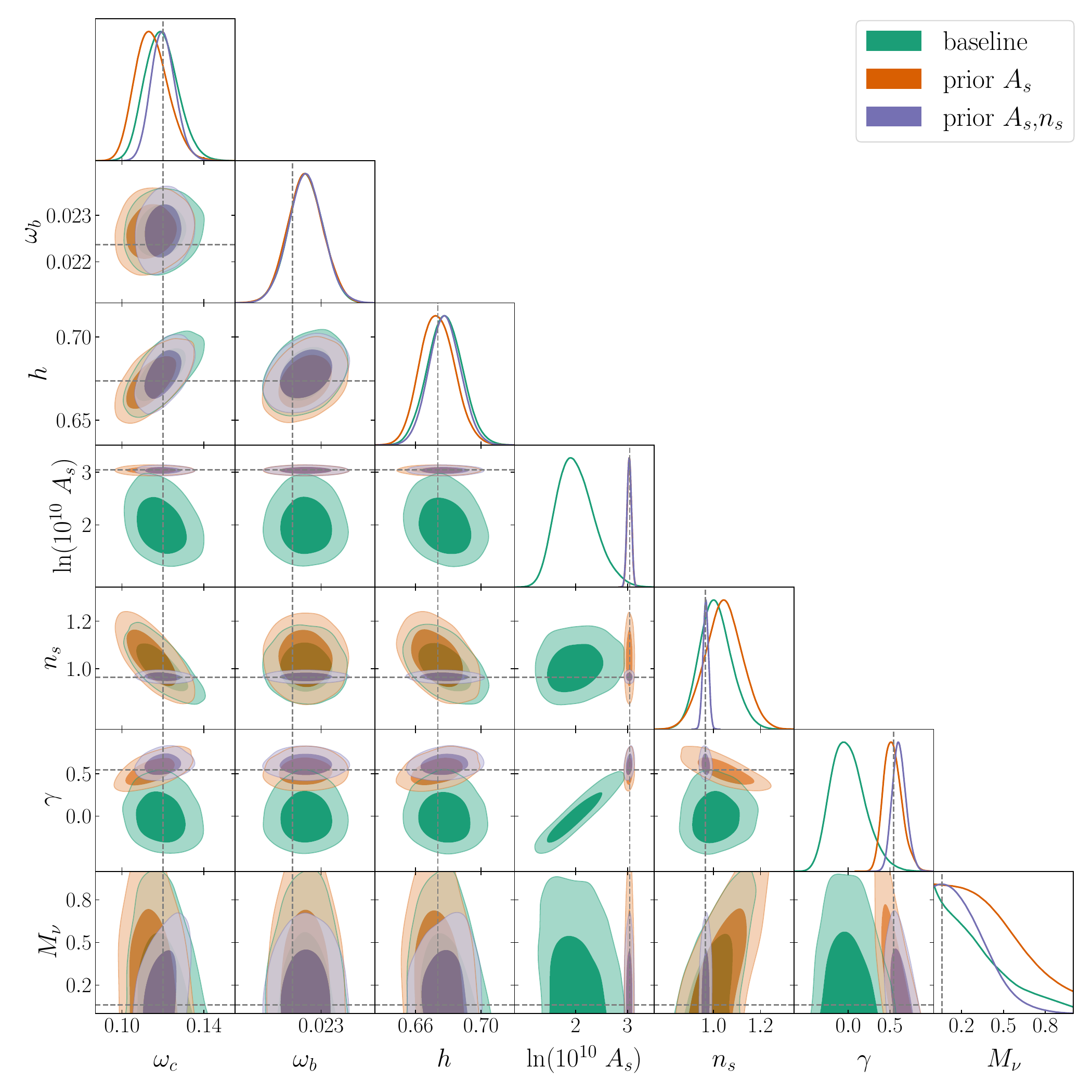}
    \caption{Marginalised posterior distribution for the cosmological
      parameters for \gnucdm{} for the three prior choices. We use $\kmax = 0.2~\hmpc$ and a flat prior $M_{\nu} \sim
      \mathcal{U}(0,1)~{\rm eV}$ for the neutrino mass. Grey dashed
      lines mark the Planck best-fit values (with the $\Lambda$CDM
      prediction for $\gamma$ and minimal neutrino mass
      $M_\nu=0.06~{\rm eV}$).}
    \label{fig:gamma-mnu-cosmo}
\end{figure}

We then open up the parameter space and allow the neutrino mass to
vary as a free parameter.  It is worth noting that we do not expect to
obtain tight constraints on the neutrino mass, as previous analyses
already highlighted the inability of BOSS data alone to do so
\cite{damico2020, colas2020, semenaite2022}; in fact, as discussed in
\cite{damico2020} the precision of BOSS data does not allow to detect
the step in the power spectrum which marks the free streaming length
of massive neutrinos, restricting detectable signatures only to changes to the overall
amplitude. Nevertheless, our goal is to study possible degeneracies
between $\gamma$ and $M_\nu$, and the impact of massive neutrinos on
the constraints on $\gamma$. We modify the theoretical model as
described in Sec.~\ref{sec:neutrinos}, and explore the same three
options for priors as for the \gcdm{} case.  The marginalised
posteriors for \gnucdm{}  are shown in Fig.~\ref{fig:gamma-mnu-cosmo},
and the 68\% c.l., mean and best-fit values for the cosmological
parameters are listed in Table~\ref{tab:gamma-mnu-bestfits}.

We first note that the baseline shows a similar shift in the two-dimensional
marginalised posteriors for $\gamma$ and $A_s$ as the one of
Fig.~\ref{fig:gamma-cosmo}. This is again due to the presence of
strong degeneracies between the parameters that control the amplitude
of the power spectrum.  However, the error on $\gamma$ is not affected
by the introduction of $M_\nu$, and in fact there seems to be no
degeneracy between the two.  This is actually due to the fact that
such degeneracy is mild compared to the others and to the size of the
error bars on BOSS data (see e.g. Fig.~\ref{fig:desi-cmb} where we
perform forecasts for Stage-IV surveys, an anti-correlation between
$\gamma$ and $M_\nu$ can clearly be seen).

When we impose Planck priors on the primordial parameters on the other
hand we can see a mild degeneracy emerge between $\gamma$ and $M_\nu$,
which leads to marginally lower values for $\gamma$ with respect to
\gcdm{}.  Concerning neutrinos, as expected the data are not very
constraining, and we only provide (somewhat large) upper limits. It is
worth commenting on the case with a CMB prior on $A_s$, where the model
seems to be picking up a large value for $M_\nu$: we argue that this
is to be ascribed to the model trying to lower the amplitude by
exploiting the degeneracy between $M_\nu$ and $n_s$.  More
specifically, in the baseline case the amplitude is mostly controlled
by $A_s$; when that is fixed by the prior, other parameters change to
compensate: mostly $\gamma$, but also the neutrino mass, resulting in a looser constraint on $M_\nu$ in that case. However, this is
only possible if the change in the shape of the power spectrum due to
massive neutrinos is compensated by a change in $n_s$. When $n_s$ is
also fixed by the Planck prior, there is less freedom in the shape and
thus the constraint on $M_\nu$ tightens again.

\begin{table}[ht!]
\footnotesize
  \centering
  \begin{tabular} { l  c  c  c}
    Parameter &  baseline & prior on $A_s$  &  prior on $A_s$,$n_s$\\
    \hline
	\hline
\multirow{2}{2em}{$\omega_c       $} & $0.1193^{+0.0072}_{-0.0088}$ & $0.1146^{+0.0070}_{-0.0090}$ & $0.1202^{+0.0054}_{-0.0063}$\\
& (0.1173)	& (0.1184) 	& (0.1195)  \\
\hline
\multirow{2}{2em}{$\omega_b       $} & $0.02266\pm 0.00038        $ & $0.02265\pm 0.00039        $ & $0.02267\pm 0.00038        $\\
& (0.02284)	& (0.02292) & (0.02264) \\
\hline
\multirow{2}{2em}{$h              $} & $0.678\pm 0.011            $ & $0.673\pm 0.011            $ & $0.6779\pm 0.010          $\\
& (0.675)	& (0.683) 	& (0.681)   \\
\hline
\multirow{2}{2em}{$\ln(10^{10}~A_s)$} & $2.004^{+0.298}_{-0.412}     $ & $3.035\pm 0.042            $ & $3.034\pm 0.042           $\\
& (2.290)	& (3.028)  	& (3.040)    \\
\hline
\multirow{2}{2em}{$n_s            $} & $1.009^{+0.061}_{-0.074}   $ & $1.043\pm 0.079            $ & $0.967\pm 0.012            $\\
& (0.977)	& (1.028) 	& (0.957)   \\
\hline
\multirow{2}{2em}{$\gamma         $} & $0.001^{+0.165}_{-0.232}      $ & $0.534^{+0.091}_{-0.12}    $ & $0.612^{+0.075}_{-0.090}   $\\
& (0.183)	& (0.526) 	& (0.61)    \\
\hline
\multirow{2}{2em}{$M_{\nu}        $} & $< 0.372                   $ & $< 0.478                   $ & $< 0.298                   $\\
& (0.038)	& (0.362) 	& (0.016)   \\
\hline
\multirow{2}{2em}{$\sigma_8       $} & $0.496^{+0.054}_{-0.108}    $ & $0.806^{+0.031}_{-0.034}   $ & $0.809\pm 0.031            $\\
& (0.550)	& (0.818)	& (0.806)   \\
    \hline
  \end{tabular}
  \caption{Mean values and 68\% c.l. values for \gnucdm{} for the three
    prior choices. We show the best-fit values in parentheses, and include
    derived constraints on $\sigma_8$.}
	\label{tab:gamma-mnu-bestfits}
\end{table}

\begin{figure}
    \includegraphics[width=0.5\textwidth]{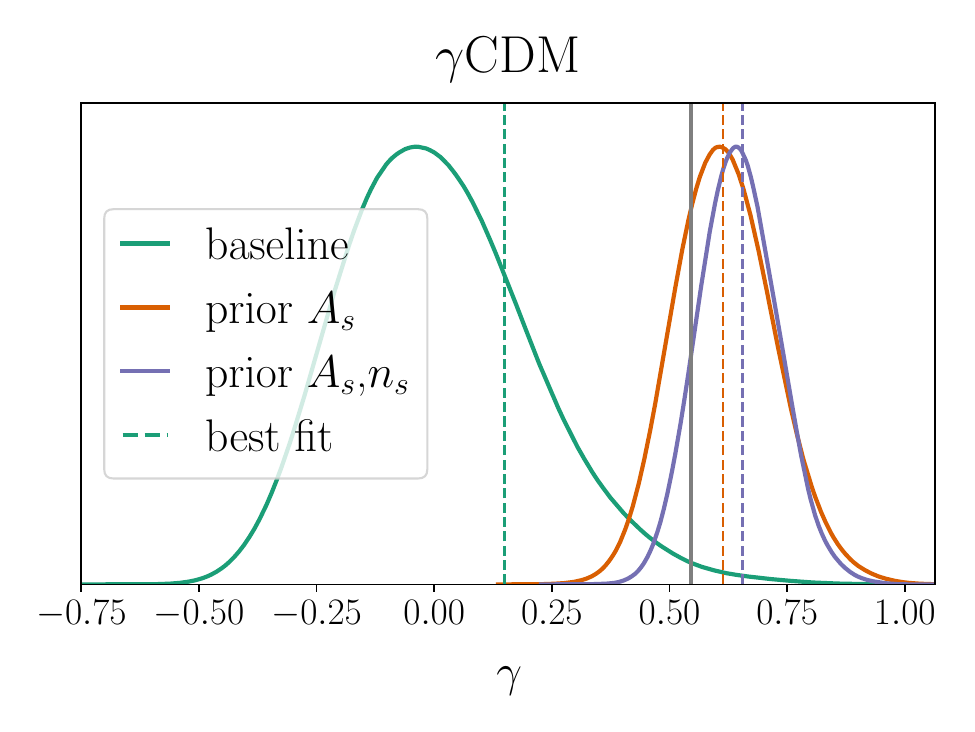}
    \includegraphics[width=0.5\textwidth]{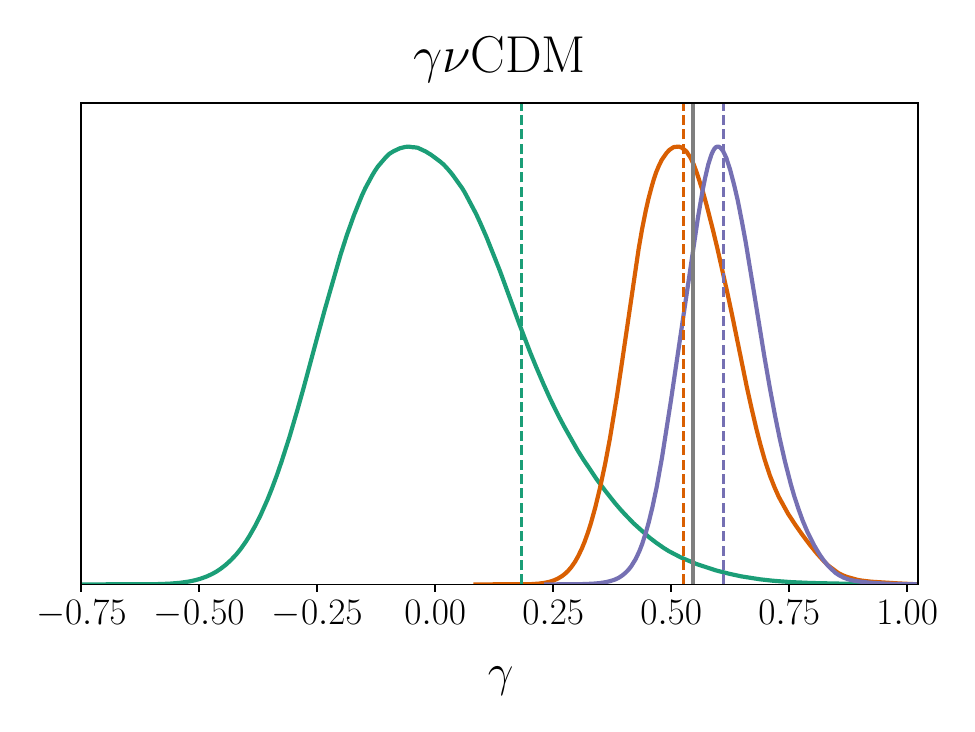}
    \caption{Marginalised one-dimensional posterior distribution for the $\gamma$
      parameter for \gcdm{} (left) and \gnucdm{} (right),
      for the three prior choices as stated in the legend. Dashed
      vertical lines mark the best-fit values corresponding to the
      maximum of the marginalised posterior distribution; the solid
      grey line corresponds to the $\Lambda$CDM prediction.}
    \label{fig:gamma-1d}
\end{figure}

In Fig.~\ref{fig:gamma-1d} we plot the one-dimensional marginalised posterior
distributions for $\gamma$ for all cases described in this section, to
allow for an easier comparison between the constraints we get for this
parameter. We mark the best-fit values with dashed lines, following the same
color scheme of Fig.~\ref{fig:gamma-cosmo} and
\ref{fig:gamma-mnu-cosmo}, and the $\Lambda$CDM prediction with solid
grey lines.  As we can see, adopting CMB priors on the primordial
parameters shifts the peak of the posterior from (projection affected)
negative values to values consistent with the $\Lambda$CDM prediction,
albeit slightly larger. Moreover, the difference between the peak of
the posterior and the best-fit for the case without CMB-based priors
is a hint of the projection which affects the marginalised posteriors.

Our results are broadly consistent with previous analyses of the BOSS
data, although an exact comparison is difficult due to different
choices in the analysis setup and datasets included. The main novelty
of this work is the use of a full-shape approach, as opposed to
template fitting adopted in the official BOSS analysis, and the
exploration of the impact of priors on the inferred
parameters. Moreover, most previous works take advantage of a joint
analysis with CMB data, which results in generally tighter
constraints. Nevertheless, we report here some of the recent
measurements of $\gamma$ that used BOSS galaxies as the main dataset
for comparison, highlighting the main differences with respect to this
work: \cite{mueller2018} found $\gamma = 0.566 \pm 0.058$ combining
BOSS DR12 galaxies with Planck CMB data and SNIa data. Additionally,
\cite{beutler2014b} provides joint constraints on $\gamma$ and the
neutrino mass $M_{\nu}$, although they used a previous data release
(DR11) and again performed a joint fit with CMB data: they found
$\gamma = 0.67 \pm 0.14$, $M_\nu = 0.25 ^{+0.13}_{-0.22}$. A joint
analysis with the galaxy bispectrum was performed in
\cite{gilmarin2017}, finding $\gamma = 0.733^{+0.068}_{-0.069}$. All
these works are official analyses from the BOSS collaboration, and
adopt a fixed template for the power spectrum multipoles.  More
recently, \cite{nguyen2023} performed a similar analysis combining
data from BOSS, DES and Planck, finding $\gamma =
0.633^{+0.025}_{-0.024}$. Also in this case the power spectrum was
fitted using a fixed template, as opposed to the full-shape fit we
perform in this work.

Overall, our strongest constraints (those where we impose CMB-based
priors on the primordial parameters) show a marginal preference for a
high value of $\gamma$, consistent for example with
\cite{nguyen2023}. However, in our case the deviation from the
$\Lambda$CDM prediction is $\sim 1\sigma$, as opposed to the $\sim
4\sigma$ discrepancy found in that work.  In general, our constraints
are weaker than those obtained in previous works. This can be
attributed to the fact that we only rely on LSS data and include CMB
information merely in terms of priors on the primordial parameters:
given the degeneracies between the additional parameters of \gcdm{}
and \gnucdm{} and the other parameters, some of which are tightly
constrained by the CMB, it is no surprise that performing a joint
analysis with Planck data can yield tighter constraints.  This can
also be approximated by including Gaussian priors derived from Planck
on all cosmological parameters, see Appendix F of \cite{chudaykin2019}
for a comparison.

Another possible way of gaining in constraining power without
resorting to CMB information would be to use higher-order statistics
such as the galaxy bispectrum.  Indeed, the tree-level bispectrum does
not depend on $b_{\Gamma_3}$, and its inclusion in the analysis can
therefore break the degeneracy between $b_{\mathcal
  G_2}$ and
$b_{\Gamma_3}$, leading to improved constraints on the
linear bias $b_1$ (see \cite{damico2020, philcox2022} for a full-shape
joint analysis of BOSS data). While the impact on cosmological
parameters is somewhat marginal in the context of $\Lambda$CDM and
Stage-III surveys, it can become significant for extended models where
there are strong degeneracies between parameters controlling the
amplitude: a tighter constraint on $b_1$ implies a more precise
measurement of $A_s$ and therefore of $\gamma$.  Some work towards
quantifying the impact of including the bispectrum for
beyond-$\Lambda$CDM models has already been done in \cite{tsedrik2023}
in the context of simulations, where a $\sim 30\%$ improvement was
found for a power spectrum and bispectrum joint analysis for an
interacting dark energy model.  We leave a proper exploration of the impact of
the bispectrum in the context of \gcdm{} and \gnucdm{} to a future
work.
\subsection{Projection effects}
\label{sec:projection-effects}
In this section we study in more details the prior volume effects
found for \gcdm{} and \gnucdm{} when no CMB priors are applied (green
contours in Fig.~\ref{fig:gamma-cosmo} and \ref{fig:gamma-mnu-cosmo},
in particular the $A_s$--$\gamma$ plane).  The presence of projection
effects is already highlighted by the shift between the peak of the
one-dimensional marginalised posterior and the best-fit point, see
Fig.~\ref{fig:gamma-1d}. To better investigate this we adopt two
approaches: first we perform a consistency check using synthetic data,
then we carry out profiling of the posterior.

\begin{figure}[h!]
    \centering
    \includegraphics[width=0.95\columnwidth]{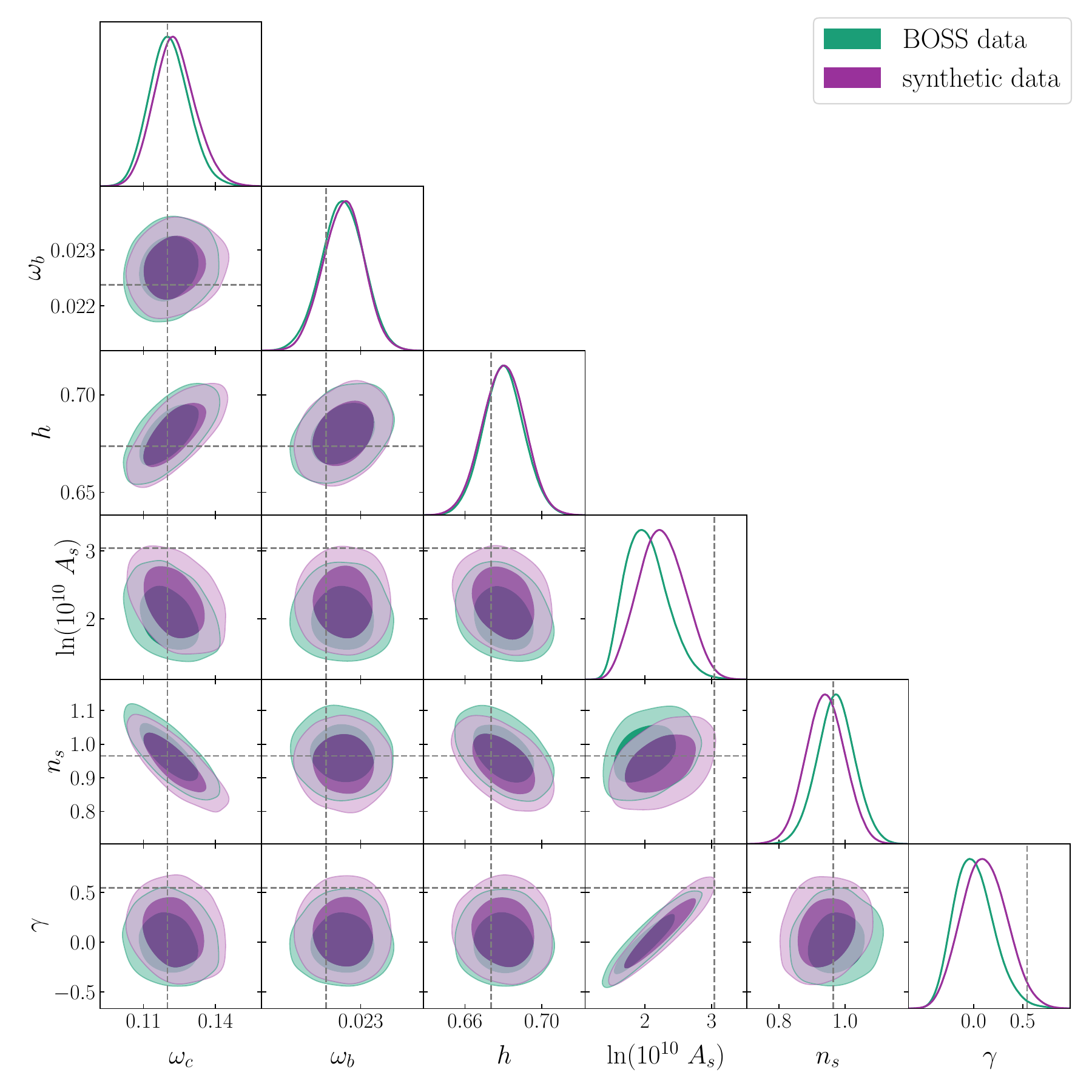}
    \caption{Posterior distribution for the cosmological parameters
      for the BOSS data (green) and synthetic data generated with the
      Planck fiducial cosmology (purple). We fix the total neutrino
      mass to $M_{\nu} = 0.06~{\rm eV}$ and use a scale-cut of $\kmax
      = 0.2~\hmpc$. Grey dashed lines mark the fiducial
      cosmology used to generate the synthetic data.}
    \label{fig:gamma-boss-synth}
\end{figure}

For the consistency check we perform the same analysis described in
Sec.~\ref{sec:boss-results} on a synthetic datavector generated with a
fiducial cosmology and set of nuisance parameters.  We note that the
ability of our pipeline to recover the cosmological parameters
correctly has been validated with N-body simulations, and we are using
the same theoretical model to generate the datavector and fit it,
which should lead to perfect agreement between the input and inferred
parameters.  We use the same covariance computed from the
`MultiDark-Patchy' mock catalogues adopted in the BOSS analysis, and
include (synthetic) BAO information. As fiducial parameters we adopt
the best-fit values of Planck \cite{planck2018cosmo}, while the
nuisance parameters are determined by maximising the BOSS likelihood with
cosmology fixed to the Planck values. Our results are plotted in
Fig.~\ref{fig:gamma-boss-synth} (purple contours and lines), where we
also include the baseline result of Fig.~\ref{fig:gamma-cosmo} for
comparison (green contours and lines).

The two-dimensional marginalised posteriors for the synthetic data
exhibit a very similar behaviour as the one we observed in the BOSS
data, with a strong shift in the posterior towards low values of $A_s$
and $\gamma$ and almost perfect overlap with the posterior obtained
from BOSS data.  As discussed above, this effect can be ascribed to
the presence of extreme degeneracies in the extended parameter space,
which lead to a highly non-Gaussian posterior.  As a consequence, the
marginalisation process can result in apparent biases in the
two-dimensional confidence regions, as discussed thoroughly in
\cite{gomezvalent2022}.\footnote{See also
Ref.~\cite{Hadzhiyska:2023wae} for further explanation of the origin
of these biases.} In that work, the author proposes to use profile
distributions to assess the impact of marginalisation. As opposed to
marginalisation, this profile likelihood (PL) approach allows one to
obtain a one-dimensional distribution by evaluating the posterior over
a scan of a parameter of interest, $\theta_i$, while maximising the
posterior with respect to all other parameters $\theta_k$, $k \neq
i$. This profile distribution is not biased by projection effects,
since it is centred on the maximum of the full
posterior. For non-Gaussian posteriors, as is the case here, the resulting distribution will be different than the usual marginalised distribution obtained via integration over the marginalised parameters. Depending on the type of non-Gaussianity, this distribution can be broader or narrower over the parameter of interest. This difference between integration and maximisation of the posterior is approximately given by the so-called Laplace term, which accounts for volume effects as explained in Ref.~\cite{Hadzhiyska:2023wae}. 

\begin{figure}[h!]
    \centering
    \includegraphics[width=\columnwidth]{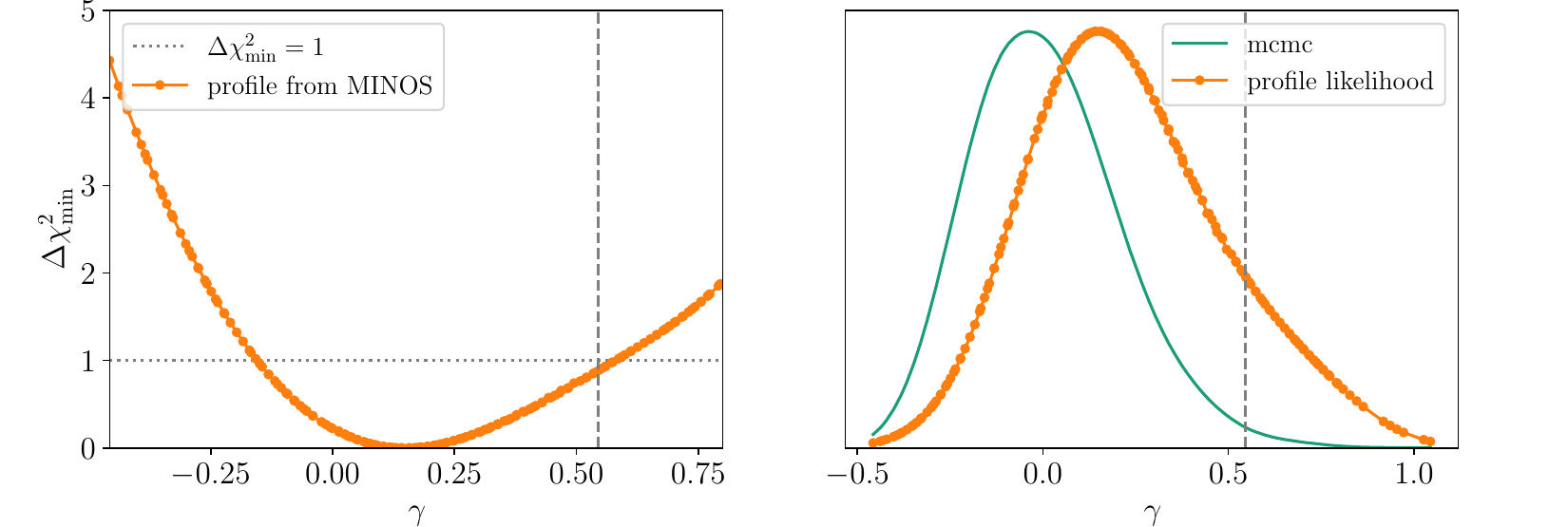}
    \caption{{\it Left:} $\Delta \chi^2_{\rm min}$ for a set of values
      for $\gamma$ obtained from profiling the likelihood with
      MINOS. The horizontal dotted line marks $\Delta \chi^2_{\rm
        min}=1$, which is used to determine confidence intervals. {\it
        Right:} comparison between the marginalised posterior obtained
      from the mcmc (green curve) and the PL (orange curve and
      datapoints). The dashed vertical line marks the $\Lambda$CDM
      prediciton.}
    \label{fig:profile-gamma}
\end{figure}

\begin{figure}[h!]
    \centering
    \includegraphics[width=\columnwidth]{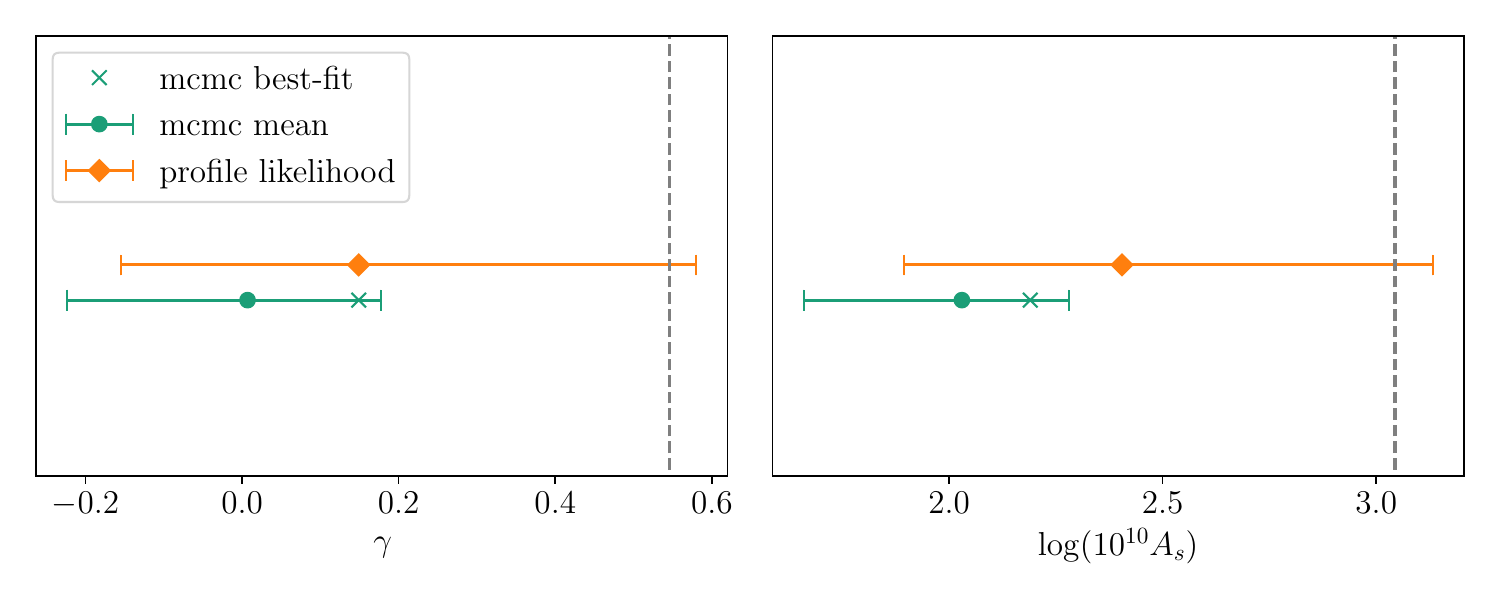}
    \caption{Comparison between the mean values and confidence
      intervals obtained from the marginalised posterior (green points
      and error bars) and the PL (orange points and error bars) for
      $\gamma$ (left) and $\ln(10^{10}~A_s)$ (right). We show the
      best-fit points obtained from the MCMC with crosses. The dashed
      vertical line marks the $\Lambda$CDM prediciton.}
    \label{fig:profile-errors}
\end{figure}

Ref.~\cite{gomezvalent2022} suggests performing this
profiling step directly on the MCMC samples. However, in addition to
resulting in noisy distributions, this is not ideal in our case, since
we sample over the analytically marginalised posterior and we
therefore expect additional projection due to that pre-marginalisation
step. We profile instead the full posterior using the MINOS algorithm
implemented in the {\tt iminuit}
package\footnote{\url{https://iminuit.readthedocs.io/en/stable/index.html}}
\cite{iminuit}. Our parameter of interest is the growth
index $\gamma$, for which we plot the PL in the left panel of
Fig.~\ref{fig:profile-gamma}, and compare to the one-dimensional
marginalised posterior from the MCMC in the right panel. Additionally,
we compare the confidence intervals obtained from the MCMC to those
obtained from the profiling in Fig.~\ref{fig:profile-errors}.

Fig.~\ref{fig:profile-gamma} shows a clear shift of the PL towards the
$\Lambda$CDM prediction with respect to the marginalised
posterior. Moreover, the confidence intervals are larger in the case
of the PL, making the results consistent with $\Lambda$CDM.  We also
note how the best-fit points from the MCMC (orange crosses in
Fig.~\ref{fig:profile-errors}) are close to the PL results, though
still somewhat lower in $A_s$. This demonstrates the presence of additional projection effects due to the analytic marginalisation of linear nuisance parameters in the posterior sampled in the MCMC. 
Overall, the use of the profile likelihood shows a more conservative result, which appears, at least in this case, to reflect better the effect of the large degeneracy between the amplitude parameters on the final one-dimensional constraint on $\gamma$. Note that should the posterior be closer to the Gaussian case, the difference between marginalising and profiling would disappear, so we expect projection effects to disappear with more constraining data.

As mentioned already in Sec~\ref{sec:results}, the presence of volume
(or projection) effects was already highlighted in a number of recent
works performing the full-shape analysis of BOSS data in the context
of beyond-$\Lambda$CDM models \cite{herold2022, simon2022,
  carrilho2023, piga2023}.  It is expected that the higher precision
of forthcoming data will mitigate this effect, as also suggested by
the forecasts we present in Sec.~\ref{sec:forecasts}. However,
projections can also be alleviated by the combination with additional
probes or the inclusion of higher order correlation functions. In
general, we recommend extreme care in the choice of priors, especially
when constraining beyond-$\Lambda$CDM models.
\subsection{Forecasts for Stage IV surveys}
\label{sec:forecasts}
In this section we present forecasts for a Stage-IV spectroscopic
galaxy survey, focusing in particular on DESI-like galaxies.
Similarly to what described in Sec.~\ref{sec:projection-effects}, we
generate and fit synthetic datavectors with the same EFTofLSS
theoretical model.  The aim is twofold: on one hand, we wish to
provide forecasted constraints on $\gamma$ and jointly on $\gamma$ and
$M_\nu$, on the other hand, we can assess the ability of the higher
precision measurement of reducing the projection effects.  Despite a
number of approximations we make in our forecasting strategy, which we
highlight below, we expect the results to give a more realistic
picture of the outcome of future surveys, as opposed to the standard
Fisher matrix approach. The latter, while extremely useful due to its
low computational cost, has been shown to be unable to capture complex
features in the likelihood, which can arise in the context of strong
degeneracies between parameters and highly non-Gaussian posteriors
(see e.g. \cite{joachimi2011, wolz2012, bellomo2020, bernal2020}).

We generate three sets of synthetic datavectors, assuming Planck
values for the cosmological parameters and simulating the three galaxy
samples that are the target of DESI: the bright galaxy sample (BGS),
luminous red galaxies (LRGs), and star-forming emission line galaxies
(ELG). We follow \cite{desi2016} and compute expected values for the
linear bias and number density of the samples, which we list in
Table~\ref{tab:desi-specs} together with the respective effective
volume.  Specifically, we assume constant values for the combination
$b_1(z)D(z)$, where $D(z)$ is the linear growth factor evaluated at
the central redshift of each bin, and use $b_{\rm LRG}(z)D(z) = 1.7$,
$b_{\rm ELG}(z)D(z) = 0.84$, and $b_{\rm BGS}(z)D(z) = 1.34$.

\begin{table}[h!]
  \centering
  \begin{tabular} { c  c  c  c  c}
    Galaxy sample & $z_{\rm eff}$ & $V_s~[{\rm Gpc}^3/h^3]$ &  $b_1$ & $\bar{n} \, [h^3/{\rm Mpc}^3]$\\
    \hline
    BGS   &  0.2  &  1.8 & 1.135 & 0.013 \\
    LRGs  &  0.8  &  12  & 2.56  & $3.2 \times 10^{-4}$\\
    ELG   &  1.2  &  17  & 1.51  & $3.7 \times 10^{-4}$\\
    \hline
  \end{tabular}
  \caption{Parameters used to generate the synthetic DESI-like
    datavectors for the forecasts. We follow \cite{desi2016} in
    computing the effective redshift $z_{\rm eff}$, the volume $V_s$ and
    number density $\bar{n}$, and determine the linear bias $b_1$
    assuming constant, sample-specific values for the combination
    $b_1(z) D(z)$ as specified in the text.}
	\label{tab:desi-specs}
\end{table}

\begin{figure}[ht!]
    \centering
    \includegraphics[width=\columnwidth]{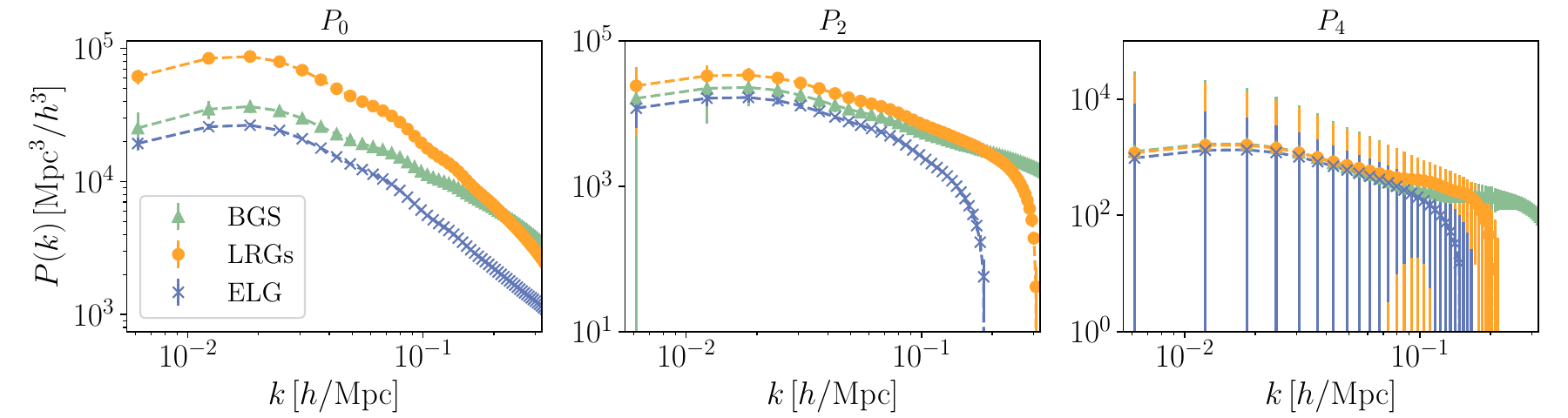}
    \caption{Power spectrum multipoles for the three galaxy samples in
      our synthetic DESI-like dataset: BGS (green, triangle markers), LRGs (orange, round markers) and ELGs (blue, crosses). The error bars are the square roots of the diagonal elements of the analytic covariance matrix.}
    \label{fig:desi-data}
\end{figure}

All redshift bins have size $\Delta z = 0.4$ and are centered on the
$z_{\rm eff}$ values listed in the table.  We compute an analytic
covariance matrix, assuming no cross correlation between the redshift
bins / samples.  Specifically, we use the
following \cite{taruya2010}:
\be
C_{\ell_1 \ell_2}(k_i, k_j) = \frac{(2\ell_1 + 1) (2\ell_2 + 1)}{N_k} \int_{-1}^1 {\rm d}\mu~\mathcal{L}_{\ell_1}(\mu_i) \mathcal{L}_{\ell_2}(\mu_j) \left[P(\bfk_i) + \frac{1}{\bar{n}} \right] \left[ P({\bfk}_j) + \frac{1}{\bar{n}} \right] \, ,
\label{eq:covariance}
\ee
where $\mu_i$ is the cosine between $\bfk_i$ and the line of sight,
$\mathcal{L}_\ell$ is the Legendre polynomial of order $\ell$,
$\bar{n}$ is the number density and $N_k = 4\pi k^2 \Delta k~V_s /
(2\pi)^3$ is the number of Fourier modes within a given $k$-bin of
size $\Delta k$, $V_s$ being the survey volume. For the power spectrum
$P(\bfk)$ in Eq.~\ref{eq:covariance} we use the Kaiser approximation
\cite{kaiser1987}:
\be
P(\bfk) = \left(b_1 + f \mu^2 \right)^2 P_L(k) \, .
\ee
As for priors, we use the same ones we adopted for the baseline BOSS
analysis of Sec.~\ref{sec:setup}, specifically we impose no prior on the
primordial parameters $A_s$ and $n_s$, both to showcase the
constraining power of LSS alone and to assess potential projection
effects in the context of Stage-IV surveys.  The resulting power
spectrum multipoles for the three samples, with corresponding
error bars, are shown in Fig.~\ref{fig:desi-data}.

\begin{figure}
    \centering
    \includegraphics[width=\columnwidth]{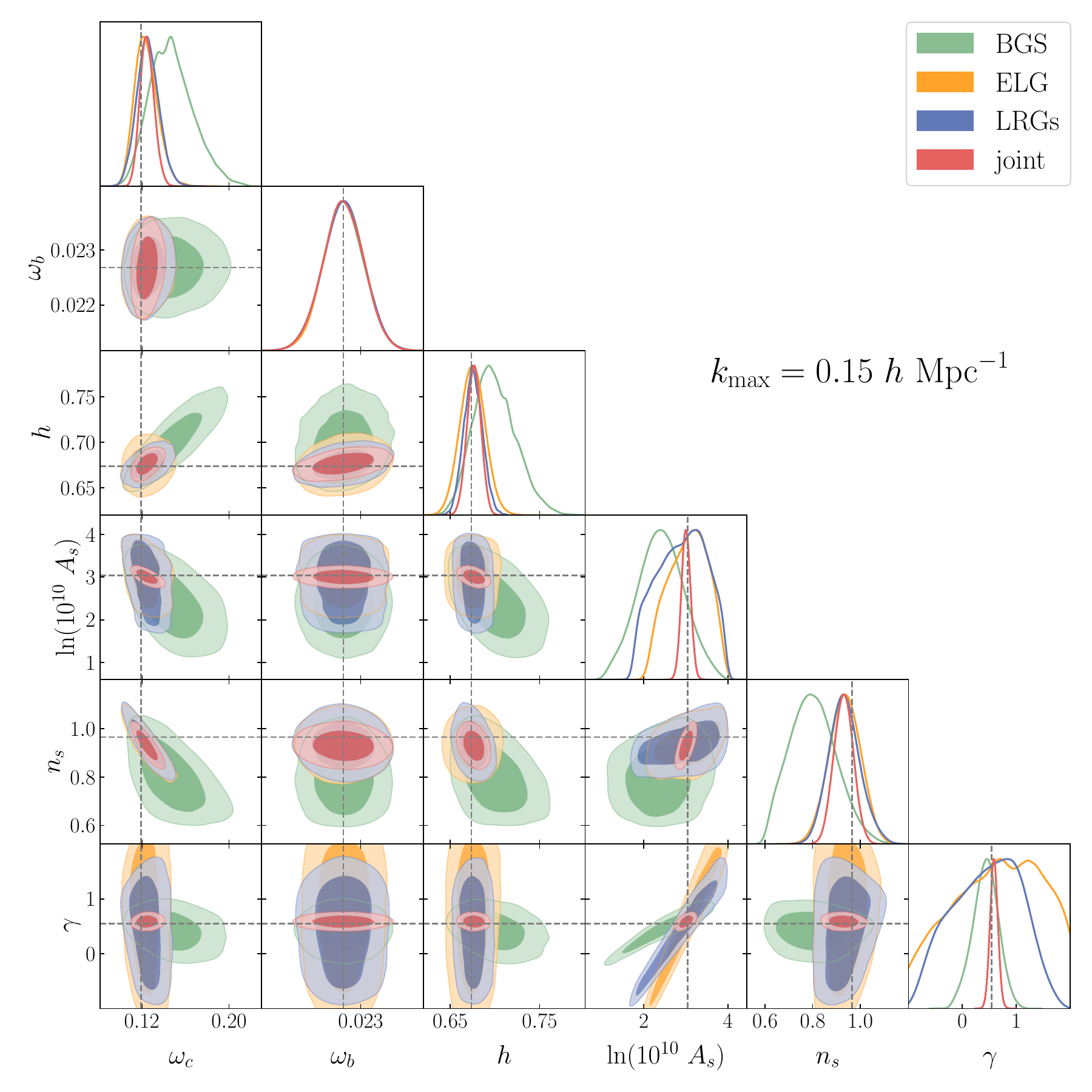}
    \caption{Marginalised posterior for \gcdm{} for the pessimistic case with $\kmax = 0.15~\hmpc$. We fit the three samples separately and jointly, as detailed in the legend. Dashed grey lines mark the fiducial values.}
    \label{fig:desi-gamma-pess}
\end{figure}

\begin{figure}
    \centering
    \includegraphics[width=\columnwidth]{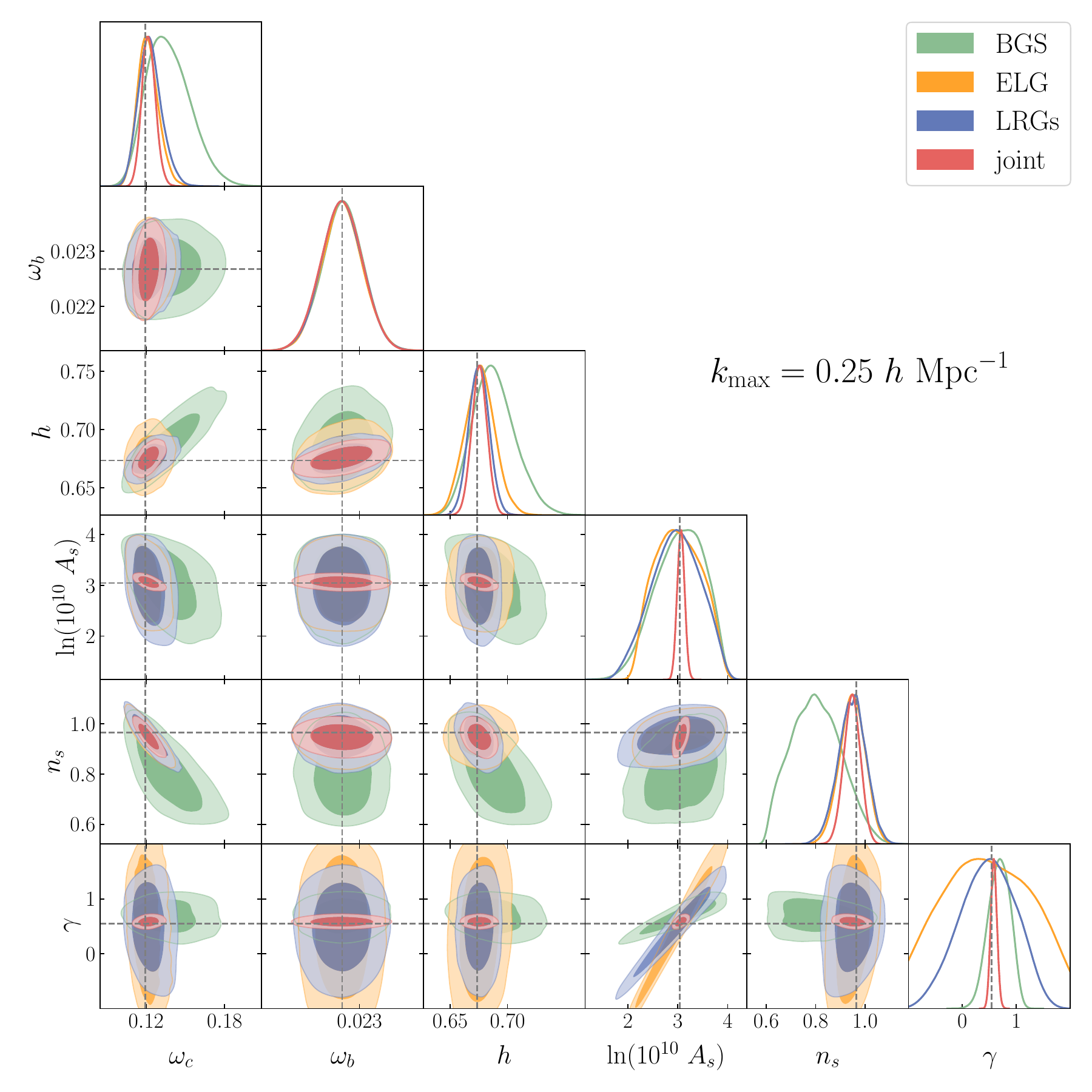}
    \caption{Marginalised posterior for \gcdm{} for the optimistic case with $\kmax = 0.25~\hmpc$. We fit the three samples separately and jointly, as detailed in the legend. Dashed grey lines mark the fiducial values.}
    \label{fig:desi-gamma-opt}
\end{figure}

\begin{figure}
    \centering
    \includegraphics[width=\columnwidth]{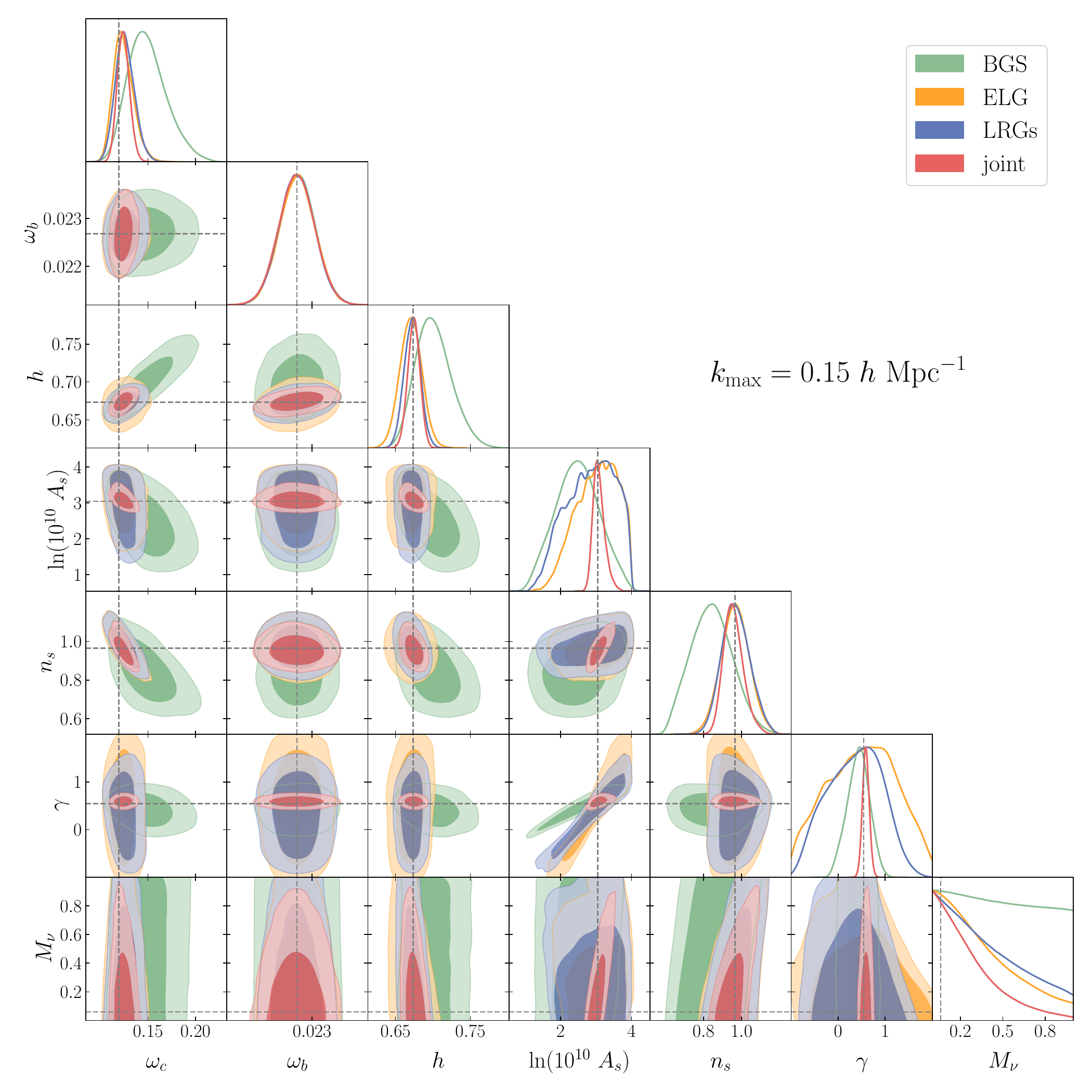}
    \caption{Marginalised posterior for \gnucdm{} for the pessimistic case with $\kmax = 0.15~\hmpc$. We fit the three samples separately and jointly, as detailed in the legend. Dashed grey lines mark the fiducial values.}
    \label{fig:desi-gamma-mnu-pess}
\end{figure}

\begin{figure}
    \centering
    \includegraphics[width=\columnwidth]{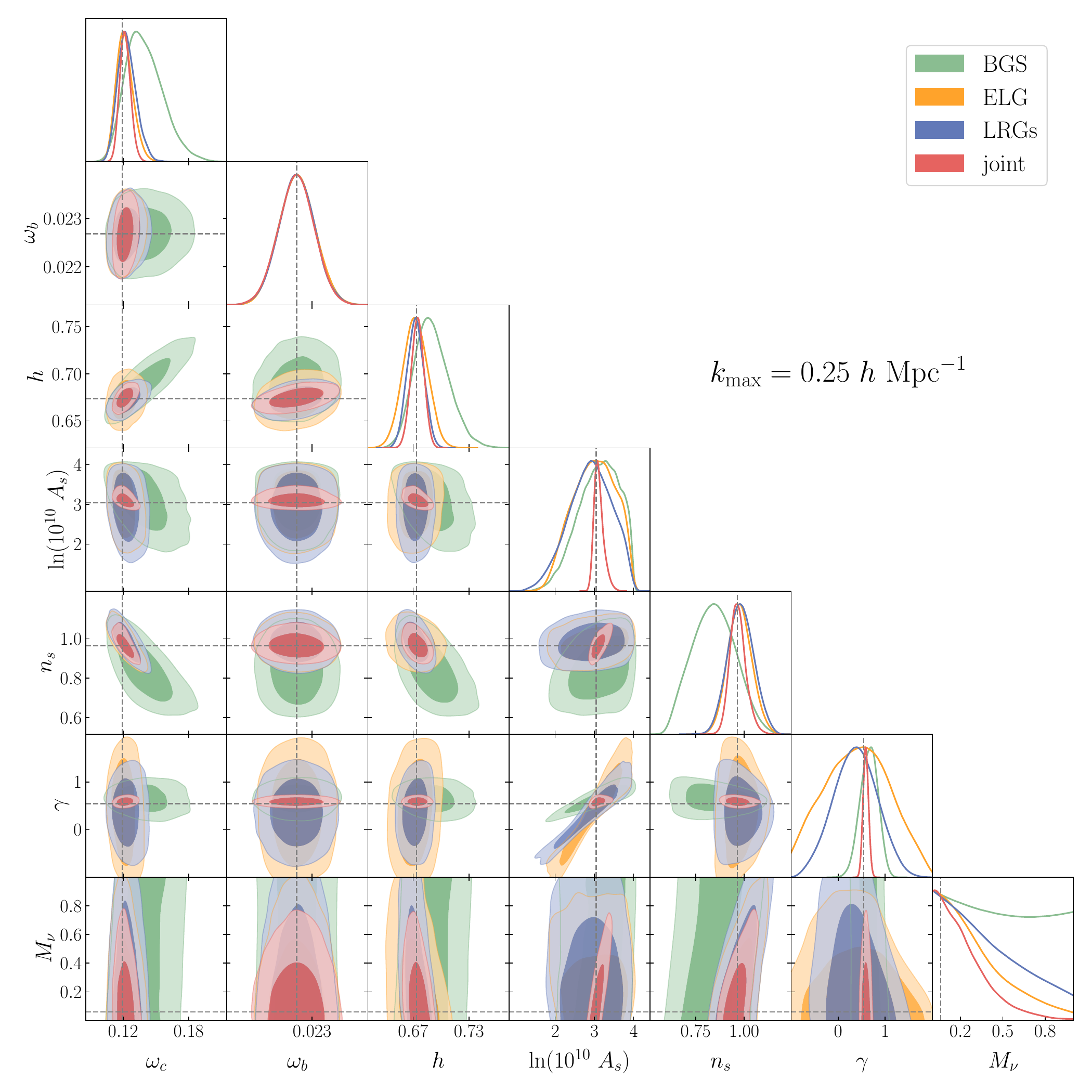}
    \caption{Marginalised posterior for \gnucdm{} for the optimistic case with $\kmax = 0.25~\hmpc$. We fit the three samples separately and jointly, as detailed in the legend. Dashed grey lines mark the fiducial values.}
    \label{fig:desi-gamma-mnu-opt}
\end{figure}

We fit the three galaxy samples separately and then combine them, and
explore two options for the scale cuts: a pessimistic one with $k_{\rm
  max} = 0.15~\hmpc$ and an optimistic one with $k_{\rm max} =
0.25~\hmpc$. Our results for \gcdm{} are shown in
Fig.~\ref{fig:desi-gamma-opt} (\ref{fig:desi-gamma-pess}) for the
optimistic (pessimistic) case, and summarised in
Table~\ref{tab:desi-gamma}. Moreover, we perform the same analysis for
\gnucdm{} and show the results in Fig.~\ref{fig:desi-gamma-mnu-opt}
(\ref{fig:desi-gamma-mnu-pess}) for the optimistic (pessimistic) case,
while the constraints are listed in Table~\ref{tab:desi-gamma-mnu}.

\begin{table}[ht!]
  \footnotesize
  \centering
  \begin{tabular} { l  c  c  c  c }
    Parameter &  BGS &  ELGs & LRGs & joint \\
    \hline
    \hline
    \multicolumn{5}{c}{Pessimistic $k_{\rm max}=0.15~\hmpc$} \\
    \hline
    \hline
    \multirow{2}{2em}{$\omega_c       $}  & $0.148^{+0.016}_{-0.023}   $ & $0.1242^{+0.0088}_{-0.012} $ & $0.1256^{+0.0095}_{-0.011} $ & $0.1250^{+0.0059}_{-0.0069}$\\
    & (0.1352) & (0.1202) & (0.1228) & (0.1212)\\
    \hline
    \multirow{2}{2em}{$\omega_b       $}  & $0.02268\pm 0.00038        $ & $0.02269\pm 0.00037        $ & $0.02268\pm 0.00038        $ & $0.02268\pm 0.00038        $\\
    & (0.02255) & (0.02230) & (0.02306) & (0.02303)\\
    \hline
    \multirow{2}{2em}{$h              $}  & $0.700^{+0.020}_{-0.027}   $ & $0.675\pm 0.014            $ & $0.676\pm 0.011            $ & $0.6765\pm 0.0078          $\\
    & (0.688) & (0.671) & (0.679) & (0.676)\\
    \hline
    \multirow{2}{2em}{$\ln(10^{10}~A_s)$} & $2.41\pm 0.55              $ & $3.06^{+0.56}_{-0.43}      $ & $2.90^{+0.71}_{-0.55}      $ & $3.00\pm 0.11              $\\
    & (2.27) & (3.26) & (3.27) & (3.04)\\
    \hline
    \multirow{2}{2em}{$n_s            $}  & $0.801^{+0.090}_{-0.11}    $ & $0.935\pm 0.066            $ & $0.933^{+0.060}_{-0.069}   $ & $0.930\pm 0.041            $\\
    & (0.752) & (0.955) & (0.950) & (0.950)\\
    \hline
    \multirow{2}{2em}{$\gamma         $}  & $0.43^{+0.25}_{-0.22}      $ & $0.64^{+1.1}_{-0.63}       $ & $0.45^{+0.76}_{-0.58}      $ & $0.587\pm 0.072            $\\
    & (0.37) & (0.94) & (0.81) & (0.57)\\
    \hline
    \multirow{2}{2em}{$\sigma_8       $}  & $0.65^{+0.11}_{-0.19}      $ & $0.84^{+0.15}_{-0.25}      $ & $0.80^{+0.15}_{-0.31}      $ & $0.804\pm 0.037            $\\
    & (0.55) & (0.903) & (0.912) & (0.809)\\
    \hline
    \hline
    \multicolumn{5}{c}{Optimistic $k_{\rm max}=0.25~\hmpc$} \\
    \hline
    \hline
    \multirow{2}{2em}{$\omega_c       $}  & $0.137^{+0.014}_{-0.019}   $ & $0.1216^{+0.0068}_{-0.0088}$ & $0.1231^{+0.0076}_{-0.0098}$ & $0.1221^{+0.0048}_{-0.0055}$\\
    & (0.1298) & (0.1175) & (0.1176) & (0.1135)\\
    \hline
    \multirow{2}{2em}{$\omega_b       $}  & $0.02269\pm 0.00038        $ & $0.02268\pm 0.00038        $ & $0.02268\pm 0.00037        $ & $0.02267\pm 0.00038        $\\
    & (0.02205) & (0.02245) & (0.02196) & (0.02139)\\
    \hline
    \multirow{2}{2em}{$h              $}  & $0.688^{+0.016}_{-0.021}   $ & $0.676\pm 0.013            $ & $0.6752\pm 0.0087          $ & $0.6755\pm 0.0067          $\\
    & (0.679) & (0.676) & (0.669) & (0.662)\\
    \hline
    \multirow{2}{2em}{$\ln(10^{10}~A_s)$} & $3.04^{+0.57}_{-0.42}      $ & $3.01^{+0.46}_{-0.51}      $ & $2.96\pm 0.48              $ & $3.063\pm 0.074            $\\
    & (2.81) & (2.91) & (3.51) & (3.10)\\
    \hline
    \multirow{2}{2em}{$n_s            $}  & $0.80\pm 0.10              $ & $0.949\pm 0.052            $ & $0.947\pm 0.056            $ & $0.947\pm 0.034            $\\
    & (0.872) & (0.975) & (0.965) & (0.982)\\
    \hline
    \multirow{2}{2em}{$\gamma         $}  & $0.68^{+0.22}_{-0.19}      $ & $0.47\pm 0.75              $ & $0.47^{+0.57}_{-0.50}      $ & $0.584\pm 0.058            $\\
    & (0.521) & (0.287) & (1.016) & (0.549)\\
    \hline
    \multirow{2}{2em}{$\sigma_8       $}  & $0.85^{+0.17}_{-0.21}      $ & $0.82^{+0.12}_{-0.25}      $ & $0.81^{+0.15}_{-0.23}      $ & $0.822\pm 0.026            $\\
    & (0.737) & (0.754) & (1.015) & (0.816)\\
    \hline
    
  \end{tabular}
  \caption{Mean values and 68\% c.l. values for \gcdm{} with fixed
    neutrino mass $M_{\nu}=0.06~{\rm eV}$ for the three DESI-like
    galaxy samples fitted separately and jointly. We show the best-fit
    values in parentheses, and include derived constaints for $\sigma_8$.}
	\label{tab:desi-gamma}
\end{table}

\begin{table}[hbtp!]
  \footnotesize
  \centering
  \begin{tabular} { l  c  c  c  c }
    Parameter &  BGS &  ELG & LRGs & joint \\
    \hline
    \hline
    \multicolumn{5}{c}{Pessimistic $k_{\rm max}=0.15~\hmpc$} \\
    \hline
    \hline
    \multirow{2}{2em}{$\omega_c       $} & $0.149^{+0.016}_{-0.024}   $ & $0.1243^{+0.0086}_{-0.012} $ & $0.1257^{+0.0091}_{-0.011} $ & $0.1242^{+0.0060}_{-0.0069}$\\
    & (0.1431 ) & (0.1171 ) & (0.1181) & (0.1212)\\
    \hline
    \multirow{2}{2em}{$\omega_b       $} & $0.02269\pm 0.00038        $ & $0.02269\pm 0.00038        $ & $0.02268\pm 0.00037        $ & $0.02268\pm 0.00038        $\\
    & (0.02322 ) & (0.02274) & (0.02261) & (0.02273)\\
    \hline
    \multirow{2}{2em}{$h              $} & $0.701^{+0.020}_{-0.027}   $ & $0.670\pm 0.015            $ & $0.672\pm 0.011            $ & $0.6745\pm 0.0081          $\\
    & (0.698 ) & (0.676 ) & (0.675) & (0.68)\\
    \hline
    \multirow{2}{2em}{$\ln(10^{10}~A_s)$} & $2.47\pm 0.61             $ & $3.02^{+0.79}_{-0.41}      $ & $2.85^{+0.96}_{-0.46}      $ & $3.08^{+0.12}_{-0.19}      $\\
    & (1.25 ) & (3.32) & (3.25) & (2.99)\\
    \hline
    \multirow{2}{2em}{$n_s            $} & $0.84\pm 0.11              $ & $0.967\pm 0.075            $ & $0.971^{+0.068}_{-0.079}   $ & $0.958^{+0.043}_{-0.058}   $\\
    & (0.816 ) & (0.969) & (0.979) & (0.95)\\
    \hline
    \multirow{2}{2em}{$\gamma         $} & $0.43\pm 0.23              $ & $0.51\pm 0.71              $ & $0.36^{+0.66}_{-0.55}      $ & $0.590\pm 0.072            $\\
    & (-0.07) & (0.04) & (0.81) & (0.563)\\
    \hline
    \multirow{2}{2em}{$M_\nu          $} & ---                          & $< 0.434                   $ & $< 0.494                   $ & $< 0.314                   $\\
    & (0.017 ) & (0.024) & (0.026) & (0.032)\\
    \hline
    \multirow{2}{2em}{$\sigma_8       $} & $0.63^{+0.11}_{-0.22}      $ & $0.79^{+0.22}_{-0.25}      $ & $0.74^{+0.19}_{-0.30}      $ & $0.801^{+0.037}_{-0.046}   $\\
    & (0.351 ) & (0.925) & (0.900) & (0.798 )\\
    \hline
    \hline
    \multicolumn{5}{c}{Optimistic $k_{\rm max}=0.25~\hmpc$} \\
    \hline
    \hline
    \multirow{2}{2em}{$\omega_c       $}  & $0.1398^{+0.0137}_{-0.0198}   $ & $0.1216^{+0.0065}_{-0.0087}$ & $0.1237^{+0.0072}_{-0.0095}$ & $0.1219^{+0.0046}_{-0.0054}$\\
    & (0.1308) & (0.1197) & (0.1208) & (0.1203 )\\
    \hline
    \multirow{2}{2em}{$\omega_b       $}  & $0.02268\pm 0.00038        $ & $0.02269\pm 0.00038        $ & $0.02268\pm 0.00037        $ & $0.02267\pm 0.00037        $\\
    & (0.02265) & (0.02278) & (0.02325) & (0.02207)\\
    \hline
    \multirow{2}{2em}{$h              $}  & $0.690^{+0.016}_{-0.021}   $ & $0.672\pm 0.013            $ & $0.673\pm 0.009          $ & $0.675\pm 0.007          $\\
    & (0.684) & (0.678) & (0.683) & (0.671)\\
    \hline
    \multirow{2}{2em}{$\ln(10^{10}~A_s)$} & $3.106^{+0.666}_{-0.393}      $ & $2.989^{+0.644}_{-0.496}      $ & $2.893^{+0.615}_{-0.516}      $ & $3.116^{+0.084}_{-0.145}    $\\
    & (2.87) & (3.12) & (3.26) & (3.07)\\
    \hline
    \multirow{2}{2em}{$n_s            $}  & $0.844\pm 0.110              $ & $0.978^{+0.058}_{-0.065}   $ & $0.983\pm 0.067            $ & $0.968^{+0.035}_{-0.045}   $\\
    & (0.877) & (0.958) & (0.979) & (0.962)\\
    \hline
    \multirow{2}{2em}{$\gamma         $}  & $0.655^{+0.202}_{-0.180}      $ & $0.405\pm 0.668              $ & $0.366\pm 0.462              $ & $0.588\pm 0.058            $\\
    & (0.539) & (0.748) & (0.795) & (0.598)\\
    \hline
    \multirow{2}{2em}{$M_\nu          $}  & ---                          & $< 0.368                   $ & $< 0.489                   $ & $< 0.270                   $\\
    & (0.214) & (0.039) & (0.052) & (0.082)\\ 
    \hline
    \multirow{2}{2em}{$\sigma_8       $}  & $0.817^{+0.181}_{-0.205}      $ & $0.776^{+0.174}_{-0.250}      $ & $0.738^{+0.166}_{-0.239}      $ & $0.815^{+0.026}_{-0.031}   $\\
    & (0.732) & (0.841) & (0.910) & (0.822)\\
    \hline
      
\end{tabular}
  \caption{Mean values and 68\% c.l. values for \gnucdm{} for the three
    DESI-like galaxy samples fitted separately and jointly. We show
    the best-fit values in parentheses, and include derived constraints
    for $\sigma_8$.}
	\label{tab:desi-gamma-mnu}
\end{table}

For the joint analysis in \gcdm{} we find $\sigma(\gamma) = 0.058 ~
(0.072)$ for the optimistic (pessimistic) case, which represents an
improvement of $\sim 85\%$ with respect to the BOSS results when no
CMB priors are imposed. For \gnucdm{} we find the same values, despite
the introduction of $M_\nu$ as a free parameter.  As for massive
neutrinos, we get $M_\nu<0.27 ~ (<0.314)$ at 68\% c.l. for the
optimistic (pessimistic) case, with a very modest improvement with
respect to the constraints obtained from BOSS data. Such a dramatic
improvement in the constraints on $\gamma$ is due to the combination
of the different samples: each sample features a slightly different
orientation of the degeneracy between $A_s$ and $\gamma$, which is
then broken when we perform the joint fit.

Focusing on the projection effects which heavily affected the BOSS analysis, we can
clearly see how they are strongly mitigated by the increased precision
of the mock data and the combination of multiple samples at different
redshifts. In fact, for the joint analysis of the three galaxy samples
the deviation between the input and best-fit values for $\gamma$ is
$\sim 0.1\sigma$, with the exception of the \gnucdm{} cosmology in the
optimistic case where the deviation is $\sim 0.4\sigma$. On the other
hand, some projection is still present in the fits for the single
samples, although we can always recover the input value at the
1$\sigma$ level.  It is also worth noting how the inclusion of more
nonlinear modes can further alleviate the projection. For example,
from the BGS sample we obtain $\gamma = 0.43^{+0.25}_{-0.22}~(0.37)$
in the pessimistic case and $\gamma = 0.68^{+0.22}_{-0.19}~(0.521)$ in
the optimistic case, with the shift between the best-fit point and the
input value reduced from $0.37\sigma$ to $0.06\sigma$.

A proper comparison to other forecasts in the literature is
complicated, because the results depend strongly on the forecasting
strategy. Nonetheless, we try to summarise previous results and
highlight the differences with this work. Overall we find weaker
constraints on $\gamma$, which can mainly be attributed to the
following reasons:
\begin{itemize}
\item most works rely on the Fisher matrix formalism, known to be unable
  to properly describe the likelihood in presence of strong
  degeneracies in the parameter space;
\item most works use a linear model for the power spectrum, with a
  significantly smaller number of nuisance parameters;
\item most works perform a joint analysis with CMB information,
  which, as discussed above, can tightly constrain some of the
  parameters which are most degenerate with $\gamma$.
\end{itemize}

Forecasts for
the $\gamma$ parameterisation for Stage-IV surveys can be found in
\cite{stril2010}. The authors use a Fisher matrix approach and a
linear model for the power spectrum, limiting their analysis at $\kmax
< 0.1~\hmpc$, and find an overly optimistic $\sigma(\gamma) = 0.0096$
for the case with fixed neutrino mass and dark energy parameter.  This
result is mainly driven by the differences in the dimensionality of
the parameter space considered: their modelling only involves six
parameters (nine when they also consider massive neutrinos and
evolving dark energy), with only one (linear) bias, and no parameter
controlling the overall amplitude of the power spectrum, whose
inclusion can undermine the ability to measure $\gamma$.

A more comprehensive collection of forecasts for future experiments
can be found in \cite{fontribera2014}, that however still relies on a
Fisher matrix approach. All forecasts presented there include Planck
CMB and are limited to a linear model for the power spectrum, and can
therefore be expected to be more constraining than our
analysis. Nonetheless, the authors find $\sigma(\gamma)=0.025$, which
is closer to our forecasted $\sigma(\gamma)=0.058$ than the results of
\cite{stril2010}.

We note that lifting some of the simplifications we make when
generating the synthetic dataset could improve our constraints. For
example, considering several redshift bins per sample can yield more
precise measurements of the cosmological parameters, including
$\gamma$ \cite{fonseca2019, viljoen2020}, but it also leads to an
increased number of nuisance parameters and increased shot-noise per
redshift bin. Moreover, some of the effects we are not accounting for
are expected to result in reduced constraining power, for example the
assumption of a Gaussian covariance, neglecting the convolution with
the survey window function and in general additional observational
systematics which can increase the size of the error bars on the
measurements \cite{wadekar2020a, wadekar2020b}.

\begin{figure}
    \centering
    \includegraphics[width=\columnwidth]{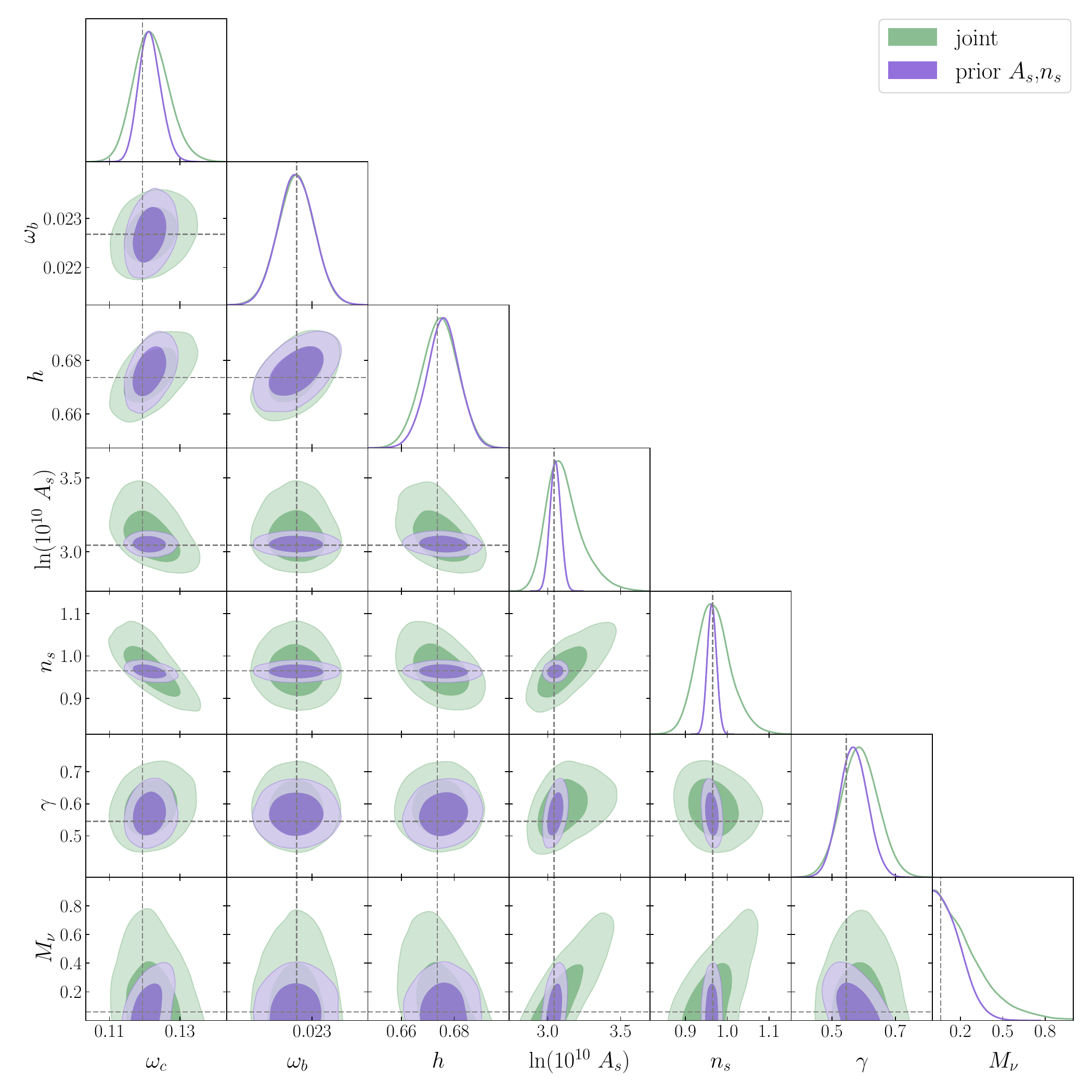}
    \caption{Comparison between the forecast when no CMB prior is
      imposed (green lines and contours) and the case with 3$\sigma$
      Planck priors on $A_s$ and $n_s$ (purple lines and contours). We use $\kmax=0.25~\hmpc$.}
    \label{fig:desi-cmb}
\end{figure}

We finally comment on our forecasts on massive neutrinos. A
measurement of the neutrino mass is one of the main goals of Stage-IV
surveys, with predicted errors well below the detection threshold
(e.g. \cite{fontribera2014, audren2013, chudaykin2019}).  In this
sense, our forecasts might seem too pessimistic, given our best case
scenario gives $M_\nu < 0.27~{\rm eV}$ at 68\% c.l. Once again, most
previous forecasts take advantage of a combined analysis with CMB data
from Planck, able to put tight constraints on the primordial
parameters $A_s$ and $n_s$. The latter are degenerate with the
neutrino mass $M_\nu$, as can be seen in
Fig.~\ref{fig:desi-gamma-mnu-opt} (or equivalently and more clearly in
the green contours of Fig.~\ref{fig:desi-cmb}): we therefore expect a
joint analysis to be extremely beneficial.  We do not perform such an
analysis here, but we can get a sense of the impact by imposing a
3$\sigma$ Planck prior on the primordial parameters, as done in
Sec.~\ref{sec:boss-results} for the BOSS analysis. We only focus on
the combination of the three redshift bins for the optimistic case
with $\kmax = 0.25 \hmpc$. Our results are shown in
Fig.~\ref{fig:desi-cmb} and Table~\ref{tab:desi-cmb}. As expected, the
CMB prior on $A_s$ and $n_s$ breaks the degeneracies and brings the
upper limit on the neutrino mass down to $M_\nu < 0.175~{\rm eV}$ at
68\% c.l., with a 35\% improvement over the case with no priors.
\begin{table}[ht!]
  \small
  \centering
  \begin{tabular} { l c c}
    Parameter &  baseline & prior on $A_s$,$n_s$\\
    \hline
    \multirow{2}{2em}{$\omega_c       $}  & $0.1219^{+0.0046}_{-0.0054}$ & $0.1215^{+0.0028}_{-0.0034}$\\
    & (0.1203 ) & (0.1211)\\
    \hline
    \multirow{2}{2em}{$\omega_b       $}  & $0.02267\pm 0.00037        $ & $0.02268\pm 0.00037        $\\
    & (0.02207) & (0.02266)\\
    \hline
    \multirow{2}{2em}{$h              $}  & $0.6746\pm 0.0069          $ & $0.6758\pm 0.0062          $\\
    & (0.6712) & (0.6767)\\
    \hline
    \multirow{2}{2em}{$\ln(10^{10}~A_s)$} & $3.116^{+0.084}_{-0.14}    $ & $3.052\pm 0.037            $\\
    & (3.074 ) & (3.016 )\\
    \hline
    \multirow{2}{2em}{$n_s            $}  & $0.968^{+0.035}_{-0.045}   $ & $0.963\pm 0.011            $\\
    & (0.962) & (0.957)\\
    \hline
    \multirow{2}{2em}{$\gamma         $}  & $0.588\pm 0.058            $ & $0.568\pm 0.045            $\\
    & (0.598) & (0.554)\\
    \hline
    \multirow{2}{2em}{$M_\nu          $}  & $< 0.270                   $ & $< 0.175                   $\\
    & (0.082) & (0.033)\\
    \hline
    \multirow{2}{2em}{$\sigma_8       $}  & $0.815^{+0.026}_{-0.031}   $ & $0.804\pm 0.020            $\\
    & (0.822) & (0.807)\\
    \hline
  \end{tabular}
  \caption{Best-fit and 68\% c.l. values for the DESI forecasts with
    3$\sigma$ Planck priors on $A_s$,$n_s$. We use
    $\kmax=0.25~\hmpc$.}
  \label{tab:desi-cmb}
\end{table}

Despite being quite large in the context of a detection of the
neutrino mass, our results are generally compatible with
\cite{noriega2022}; there, the authors performed a validation against
N-body simulations of a similar 1-loop model for the power spectrum as
the one adopted here.  For a $25~{\rm Gpc}^3/h^3$ survey, similar in
volume to our combined DESI-like mocks, they found $M_\nu < 0.49~{\rm
  eV}$ at 68\% c.l. and conclude that 2-point summary statistics from spectroscopic clustering alone might
not be sufficient detect the neutrino mass.  Beside the combination
with CMB data, we expect the combination with higher order statistics
such as the bispectrum to bring significant improvements
\cite{hahn2020, hahn2021}.  We leave the study of the impact of the
bispectrum on the determination of the neutrino mass scale to a future
work. Further improvement should come from using photometric probes, namely cosmic shear and galaxy-galaxy lensing \citep{euclidforecast2020}.

%
\section{Conclusions}
\label{sec:conclusion}
We presented the full-shape analysis of the power spectrum multipoles
measured from BOSS DR12 galaxies with the inclusion of
post-reconstruction BAO data.  We used the windowless measurements of
\cite{philcox2022} combined with BAO measurements obtained from a
variety of data-sets \cite{philcox2022, beutler2011, ross2015,
  desbourboux2020}.  We provided constraints on a single parameter
extension of $\Lambda$CDM which allows for deviations from the
standard scenario in the growth functions, the so-called `growth
index' (or $\gamma$) parameterisation. We also explored the case with
the total neutrino mass as a free parameter, and provided joint
constraints for $\gamma$ and $M_\nu$. Our theoretical model for the
power spectrum is based on the EFTofLSS and takes advantage of the
{\tt bacco} emulator for the linear power spectrum and the \texttt{FAST-PT}
algorithm for a fast evaluation of the likelihood.

We explored different options for the priors on the parameters that
determine the primordial power spectrum, $A_s$ and $n_s$, finding that
strong degeneracies in the extended parameter space result in large
projection effects, especially concerning the parameters which
regulate the amplitude of the power spectrum.  For this reason, we
applied a 3$\sigma$ Planck prior on $A_s$ and on both $A_s$ and $n_s$,
and obtained $\gamma = 0.647 \pm 0.085$ and $\gamma =
0.612^{+0.075}_{-0.090}$, respectively, $\sim 1\sigma$ consistent with the
$\Lambda$CDM prediction $\gamma = 0.55$. For neutrinos we found
$M_\nu < 0.478~{\rm eV}$ (Planck prior on $A_s$) and $M_\nu <
0.298~{\rm eV}$ (Planck prior on $A_s$, $n_s$), consistent with previous
EFTofLSS-based studies of the BOSS dataset that did not perform a full
joint analysis with CMB data \cite{damico2020, colas2020}.

To assess the presence of projection effects in the case when no priors
are imposed on the primordial parameters we generated synthetic
datavectors with a fiducial set of cosmological and nuisance
parameters, and then fitted them using the same numerical covariance used
for the BOSS analysis. We found a similar shift of the posterior in the
$A_s$--$\gamma$ plane as the one found in our baseline BOSS
analysis. Additionally, we performed a profile likelihood analysis, finding the maximum of the PL to be closer to the $\Lambda$CDM prediction than the peak of the marginalised posterior, and the confidence intervals derived from the PL to be larger and consistent with $\gamma=0.55$ at 68\% c.l.. 

Overall, we find the EFTofLSS model complemented with Planck priors to
provide constraints on $\gamma$ that are only $\sim 30\%$ larger than
the official BOSS analysis \cite{mueller2018}, despite the fact that
we did not perform a full joint analysis with CMB data. Additionally,
our theoretical model features a significantly larger number of
parameters, which is expected to degrade the constraints to some
extent. On the other hand, the full-shape approach allows to provide
direct constraints on the cosmological parameters.

In the second part of this work we presented forecasts for a Stage-IV
spectroscopic survey, focusing on a DESI-like galaxy sample. We
generated three synthetic datavectors at three redshifts using
different values for the nuisance parameters, to match the expected
BGS, LRGs and ELGs samples that are the target of DESI.  We computed
Gaussian covariance matrices neglecting the cross-correlation between
samples, and adopted two different scale-cuts: $\kmax = 0.15~\hmpc$
(pessimistic) and $\kmax = 0.25~\hmpc$ (optimistic).  We performed
separate fits for each sample and then combined them in a joint
analysis, finding the combination to provide significantly tighter
constraints, yielding $\sigma(\gamma) = 0.058 ~ (0.072)$ in the
optimistic (pessimistic) case, with a $\sim 85\%$ improvement with
respect to our baseline BOSS analysis without CMB-based priors.
Concerning neutrinos, we found that the improvement with respect to
Stage-III constraints is only marginal, $M_\nu < 0.27~{\rm eV}$
($M_\nu < 0.314~{\rm eV}$) in the optimistic (pessimistic) case at
68\% c.l..

In order to reduce the error bars on $\gamma$ and $M_\nu$, but also to
keep the projection effects under control, we advocate the combination
with additional observables.  For example, a joint analysis with
Planck data (or the inclusion of CMB-based priors, at least on the
primordial parameters), can have a significant impact on the
constraints.  We explored this by performing a fit of the synthetic
DESI-like data where we adopted 3$\sigma$ Planck priors on $A_s$ and
$n_s$. We obtained a $\sim 20\%$ improvement in the measurement of
$\gamma$, and $\sim 35\%$ for $M_\nu$ with respect to the case with no
priors.  Similarly, a combination with weak lensing can alleviate the degeneracies
and yield tighter constraints \citep{euclidforecast2020}. However, an optimal exploitation of
galaxy clustering data alone can already go in this direction: as
shown in \cite{tsedrik2023}, the ability of the bispectrum to better
determine the bias parameters can prove crucial in the context of
extended models. We leave the study of the impact of the bispectrum
for \gcdm{} and \gnucdm{} to a future work.

\appendix
\section{Full contourplots}
\label{app:full-contours}
\begin{figure}
    \centering
    \includegraphics[width=\columnwidth]{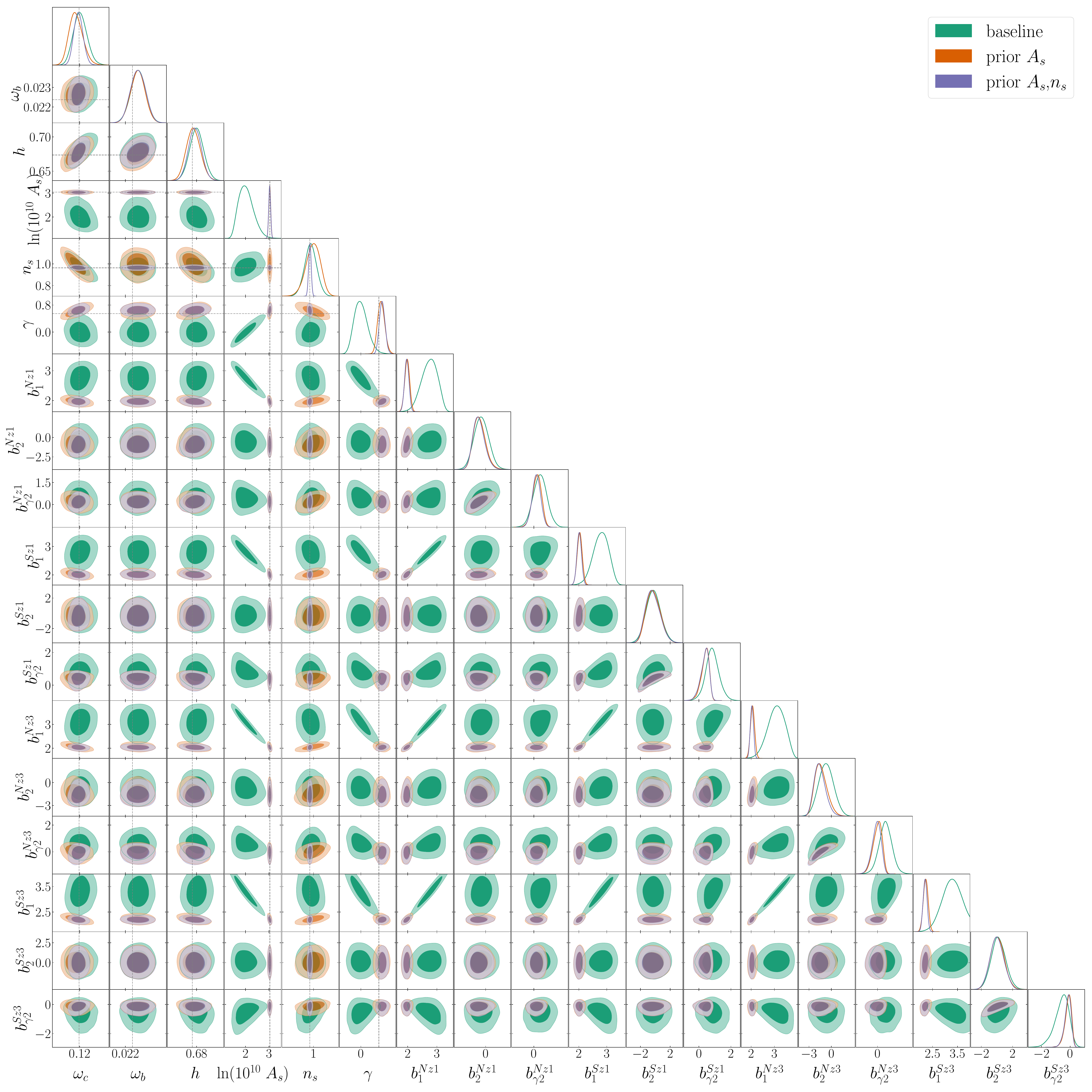}
    \caption{Two-dimensional marginalised posterior for all parameters
      for the BOSS analysis for \gcdm{} and $\kmax = 0.2~\hmpc$
      (see Fig.~\ref{fig:gamma-cosmo}). We use the same color scheme
      as the main text to show the three prior choices. Dashed grey lines mark the
      Planck best-fit values ($\Lambda$CDM prediction for $\gamma$).}
    \label{fig:gamma-full}
\end{figure}

\begin{figure}
    \centering
    \includegraphics[width=\columnwidth]{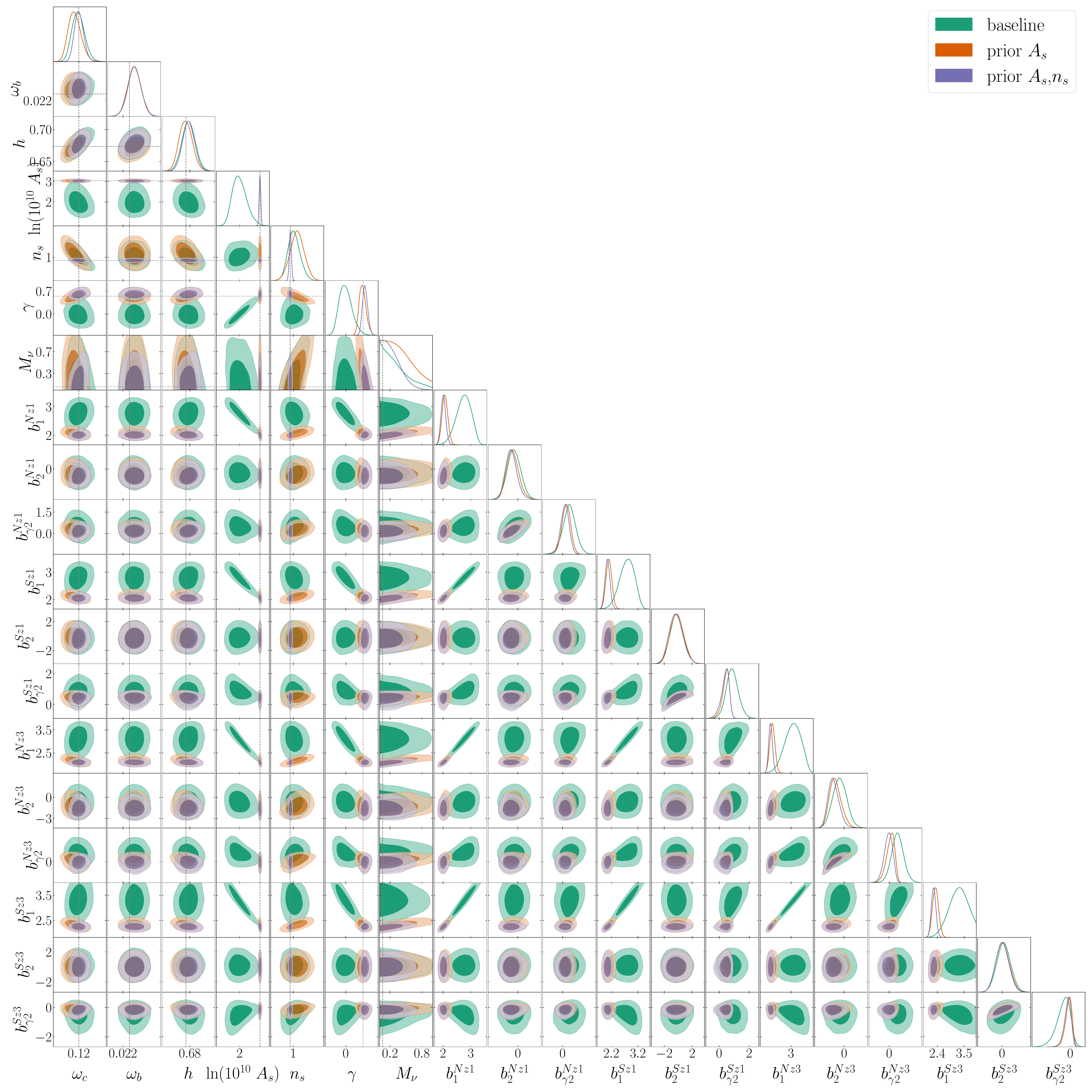}
    \caption{Two-dimensional marginalised posterior for all parameters
      for the BOSS analysis with \gnucdm{} and $\kmax = 0.2~\hmpc$
      (see Fig.~\ref{fig:gamma-mnu-cosmo}). We use the same color
      scheme as the main text to show the three prior choices. Dashed grey
      lines mark the Planck best-fit values (with the $\Lambda$CDM
      prediction for $\gamma$ and minimal neutrino mass
      $M_\nu=0.06~{\rm eV}$).}
    \label{fig:gamma-mnu-full}
\end{figure}

We show here the two-dimensional marginalised posteriors for the full
parameter space sampled in the BOSS analysis presented in
Sec.~\ref{sec:boss-results}. The results for \gcdm{} are shown in Fig.~\ref{fig:gamma-full}, corresponding to the constraints on cosmological parameters shown in Fig.~\ref{fig:gamma-cosmo}, while \gnucdm{} results are plotted in Fig.~\ref{fig:gamma-mnu-full}, corresponding to the constraints on cosmolofical parameters shown in Fig.~\ref{fig:gamma-mnu-cosmo}. The additional
model parameters not shown are the ones we analytically marginalise
over, namely $b_{\Gamma_3}$, the EFT counterterm parameters and the noise parameters. We follow the same color scheme as the main text for the three prior choices, and mark Planck best-fit values with dashed grey lines.

\section{Massive neutrinos with fixed \texorpdfstring{$\gamma$}{}}
\label{app:massive-neutrinos}
\begin{figure}
  \centering
  \includegraphics[width=\columnwidth]{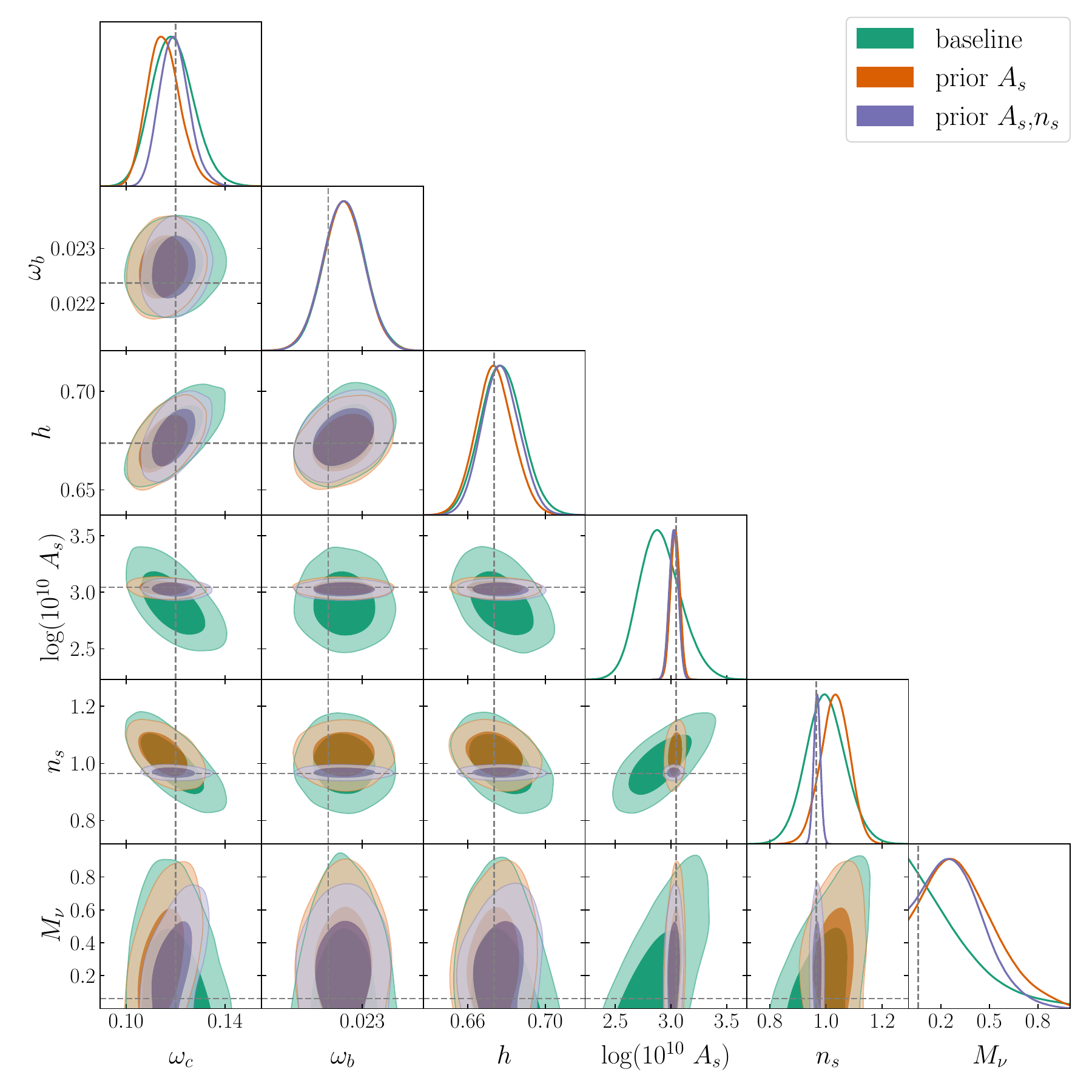}
  \caption{Marginalised posterior distribution for the cosmological
    parameters for $\Lambda$CDM with massive neutrinos and the
    three prior choices, as detailed in the legend. We fit all three
    multipoles and use $\kmax = 0.2~\hmpc$. Grey dashed
    lines mark the Planck best-fit values.}
  \label{fig:mnu-cosmo}
\end{figure}

We perform a complementary analysis of the BOSS data with free
neutrino mass but with the $\gamma$ parameter fixed to its
$\Lambda$CDM value, $\gamma=0.55$.  The two-dimensional marginalised
posteriors are shown in Fig.~\ref{fig:mnu-cosmo}, and the best-fits
are summarised in Table~\ref{tab:mnu-bestfits}.  We impose a wide,
flat prior on the neutrino mass $\mathcal{U}(0,1)~{\rm eV}$, covering
the parameter space allowed by the emulator for the linear power spectrum.  We explore again
the three options for the priors on the primordial parameters, and
obtain the following constraints for the total neutrino mass: $M_\nu <
0.319$, best-fit 0.137 (baseline), $M_\nu = 0.34^{+0.13}_{-0.28}$,
best-fit 0.141 (prior on $A_s$), $M_\nu = 0.29^{+0.12}_{-0.24}$,
best-fit 0.111 (prior on $A_s$ and $n_s$), where values are expressed
in eV.
Additionally, for the baseline analysis we find a higher value of the
amplitude: $\log(10^{10}~A_s) = 2.90^{+0.17}_{-0.21}$, as opposed to
$\log(10^{10}~A_s) = 2.67 \pm 0.17$ found in the $\Lambda$CDM analysis
of \cite{carrilho2023} (that used the same pipeline and priors adopted
here).

Comparing to previous constraints on the total neutrino mass from the
full-shape analysis of BOSS data, we find our results to be in
excellent agreement with \cite{colas2020}.  There, the authors perform
a similar EFT-based analysis of $\Lambda$CDM + massive neutrinos, also
finding that having the neutrino mass as a free parameter allows for a
slightly larger value of $A_s$, closer to the Planck best-fit value,
which is then balanced by a peak in the posterior at
$M_\nu=0.29^{+0.31}_{-0.18}~{\rm eV}$.  The same EFTofLSS model
adopted in this work was also used in \cite{ivanov2020a} to constrain
the total neutrino mass from BOSS data.  However, in that work the fit
is performed jointly with CMB data from Planck, which results in
significantly tighter constraints: $M_\nu < 0.16~{\rm eV}$ at 95\%
c.l..
\begin{table}[h!]
  \small
  \centering
  \begin{tabular} { l  c  c  c}
    Parameter &  baseline & prior on $A_s$ & prior on $A_s$, $n_s$\\
    \hline
	\hline
    \multirow{2}{2em}{$\omega_c       $} & $0.1189^{+0.0076}_{-0.0090}$ & $0.1153^{+0.0058}_{-0.0072}$ & $0.1195^{+0.0053}_{-0.0063}$\\
    & (0.1147) & (0.1099) & (0.1132)\\
    \hline
    \multirow{2}{2em}{$\omega_b       $} & $0.02267\pm 0.00038        $ & $0.02265\pm 0.00038        $ & $0.02266\pm 0.00038        $\\
    & (0.02237) & (0.02241) & (0.02260)\\
    \hline
    \multirow{2}{2em}{$h              $} & $0.677\pm 0.011            $ & $0.6739\pm 0.0097          $ & $0.6770\pm 0.0096          $\\
    & (0.675) & (0.676) & (0.676)\\
    \hline
    \multirow{2}{2em}{$\log(10^{10}~A_s)$} & $2.90^{+0.17}_{-0.21}    $ & $3.037\pm 0.041            $ & $3.024\pm 0.040            $\\
    & (3.00) & (3.07) & (3.04)\\
    \hline
    \multirow{2}{2em}{$n_s            $} & $0.998\pm 0.071            $ & $1.032^{+0.054}_{-0.048}   $ & $0.968\pm 0.012            $\\
    & (0.997) & (1.048 ) & (0.9691)\\
    \hline
    \multirow{2}{2em}{$M_{\nu}        $} & $< 0.319                   $ & $0.34^{+0.13}_{-0.28}      $ & $0.29^{+0.12}_{-0.24}      $\\
    & (0.137) & (0.141) & (0.111)\\
    \hline
    \multirow{2}{2em}{$\sigma_8       $} & $0.763^{+0.049}_{-0.081}   $ & $0.807^{+0.029}_{-0.033}   $ & $0.802^{+0.027}_{-0.032}   $\\
    & (0.783) & (0.802) & (0.779)\\
    \hline
  \end{tabular}
  \caption{Best-fit and 68\% c.l. values for the analysis with free
    neutrino mass and $\gamma=0.55$ for the three prior choices. We
    also derive best-fit values for $\sigma_8$.}
  \label{tab:mnu-bestfits}
\end{table}

\acknowledgments We thank Andrea Oddo and Emiliano Sefusatti for the effort in building the initial likelihood code. We also thank Elisabeth Krause and Catherine Heymans for useful discussions. We acknowledge use of the Cuillin computing cluster
of the Royal Observatory, University of Edinburgh.  AP is a UK
Research and Innovation Future Leaders Fellow [grant MR/S016066/2]. CM
and PC's research for this project was supported by a UK Research and Innovation Future
Leaders Fellowship [grant MR/S016066/2].  CM's work is supported by
the Fondazione ICSC, Spoke 3 Astrophysics and Cosmos Observations,
National Recovery and Resilience Plan (Piano Nazionale di Ripresa e
Resilienza, PNRR) Project ID CN\_00000013 ``Italian Research Center on
High-Performance Computing, Big Data and Quantum Computing'' funded by
MUR Missione 4 Componente 2 Investimento 1.4: Potenziamento strutture
di ricerca e creazione di "campioni nazionali di R\&S (M4C2-19 )" -
Next Generation EU (NGEU).  MT's research is supported by a doctoral studentship in the School of Physics and Astronomy, University of Edinburgh. PC's research is supported by grant RF/ERE/221061. For the purpose of open access, the author
has applied a Creative Commons Attribution (CC BY) licence to any
Author Accepted Manuscript version arising from this submission. This work made use of publicly available software, in addition to the ones cited in the main text, we acknowledge use of the {\tt numpy} \cite{numpy}, {\tt scipy} \cite{scipy}, {\tt matplotlib} \cite{matplotlib} and {\tt getdist} \cite{getdist} Python packages.

\bibliographystyle{JHEP}
\bibliography{biblio.bib}

\end{document}